\documentclass{aa}  

\usepackage{graphicx}
\usepackage{txfonts}
\usepackage{soul}

\usepackage{float}
\usepackage{xcolor}

\newcommand*{\kms}{km\,s$^{-1}$}
\newcommand*{\Elow}{E$_{\mathrm{low}}$}
\newcommand*{\loggf}{$\log gf$}
\newcommand*{\rv}{$v_{\mathrm{rad}}$}

\usepackage{hyperref}

\begin{document}

   \title{Atmospheric dynamics of the hypergiant RW\,Cep during the Great~Dimming}

   \author{A. Kasikov\inst{\ref{to}}
          \and
          I. Kolka\inst{\ref{to}}
          \and
          A. Aret\inst{\ref{to}}
          \and
          T. Eenmäe\inst{\ref{to}}
          \and
          S. P. D. Borthakur\inst{\ref{to},\ref{sri},\ref{graz}}
          \and
          V. Checha\inst{\ref{to}}
          \and
          V. Mitrokhina\inst{\ref{to}}
          \and
          S. Yang\inst{\ref{victoria}}
          }

   \institute{Tartu Observatory, University of Tartu, Observatooriumi 1, Tõravere, 61602, Estonia\label{to}\\
              \email{anni.kasikov@ut.ee}
              \and
              Space Research Institute, Austrian Academy of Sciences, Schmiedlstrasse 6, 8042, Graz, Austria\label{sri}
              \and
              Institute for Theoretical and Computation Physics, Graz University of Technology, Petersgasse 16, 8010 Graz, Austria\label{graz}
              \and
              University of Victoria, Department of Physics and Astronomy, Victoria, B.C., Canada\label{victoria}
             }

   \date{Received ; accepted }

  \abstract 
   {The hypergiant RW\,Cep is one of the largest stars in our galaxy. The evolution and mass loss of such stars has profound effects on their surrounding regions and the galaxy as a whole. Between 2020 and 2024, RW\,Cep experienced a historic mass-loss event known as the Great Dimming. }
   {This study provides a spectroscopic analysis of RW\,Cep during the Great Dimming. We examine its atmospheric dynamics and place it in the context of the star's variability behaviour since the early 2000s.}
   {We conducted high-cadence spectroscopic observations of RW\,Cep during the dimming event using the Tartu Observatory 1.5-meter telescope and the Nordic Optical Telescope. We analysed the atmospheric dynamics by measuring the radial velocities and line depths of \ion{Fe}{I} and other spectral lines. }
   {The radial velocities of the \ion{Fe}{I} lines reveal a vertical velocity gradient of {10--20~\kms} in the atmosphere, correlating with the strength of the spectral lines. Stronger lines, formed in higher atmospheric layers, have higher radial velocities. We measured the systemic velocity at $-50.3$~\kms. During the dimming, radial velocities were affected by additional emission from the ejected gas, which was blue-shifted relative to the absorption lines. Post-dimming, we observed large-scale atmospheric motions with amplitude $\sim25$~\kms. Strong resonance lines of \ion{Ba}{II}, \ion{K}{I}, \ion{Na}{I} and \ion{Ca}{I} showed stable central emission components at $-56$~\kms, likely of circumstellar origin. }
   {}


   \maketitle

\section{Introduction}

\object{RW Cep} has been included in various studies of variable and massive stars during the second half of the 20th century. Many of them were not specifically focusing on RW\,Cep, although several authors wrote about the exceptional spectral features of the star (e.g. \cite{merrill_complex_1956,gahm_spectral_1972}). RW\,Cep gained significantly more attention after  W.~Vollmann and C.~Sigismondi published a notice in the Astronomer's Telegram about the star undergoing a historic dimming event \citep{ATel15800}. In this paper, we will take a spectroscopic look at this historic event by using high temporal cadence observations taken during and after the dimming minimum, and we put this event in the context of previous variability of the star based on archival data from the beginning of the 2000s. 

RW\,Cep has been categorised as an irregular variable at least since 1933, with variability reported by different observers to be near 0.5\,mag or up to 1.2\,mag \citep{rajchl_observations_1933}. \cite{keenan_luminosities_1942} measured the visual magnitude of RW\,Cep in December 1941 to be variable 6.8--7.5\,mag with a luminosity class Ia-0 and spectral class M0. However, the determination of the spectral class of RW\,Cep has been a subject of much discussion: soon afterwards, it was classified as a G8 star \citep{morgan_revised_1950}, as a K0 type star \citep{gahm_spectral_1972,humphreys_studies_1978}, a K2 class star \citep{keenan_perkins_1989}, and back to G-type supergiant \citep{josselin_atmosphere_2003,josselin_atmospheric_2007}. The uncertainty in determining the spectral class came from the extreme range in excitation of lines in the spectrum of the star \citep{morgan_revised_1950} or, as \cite{Leadbeater2023} suggests, due to different criteria or spectral regions used by different authors, rather than significant variability in the spectrum of the star. Nevertheless, the authors can agree on one thing: RW\,Cep is one of the most luminous stars known in our galaxy  \citep{keenan_luminosities_1942,humphreys_studies_1978}.

The spectral features of RW\,Cep have always been rather unusual. The enhanced strength of the spectral lines and their irregular shapes have warranted a few more detailed studies \citep[e.g.][]{merrill_complex_1956,josselin_atmospheric_2007}. \cite{gahm_spectral_1972} commented on the abnormally broad and strong lines in the spectrum of RW\,Cep, emphasising that the star is unique among late-type supergiants and the only star with similar characteristics is $\rho$\,Cas. Parallels between RW\,Cep and other yellow hypergiants (YHGs) have also been drawn by \cite{humphreys_studies_1978}, who wrote that its spectrum resembles two other known YHGs: HR\,5171a and HR\,8752 (V509\,Cas). \cite{lobel_high-resolution_2003} also note the spectral similarities between $\rho$ Cas and RW\,Cep in late 1998. RW\,Cep displayed broadened lines characteristic of hypergiants and doubled line cores in some low-energy photospheric absorption lines.

Infrared excess has been observed in the spectral energy distribution. It has been characterised by 150--250~K \citep{jones_recent_2023} or 300~K \citep{stickland_iras_1985} black body emission. \cite{jones_recent_2023} determined that a dusty shell is located at 1000--1400~au from the star and it is compatible with a previous mass-loss ejection $\sim$95--140 years ago. Before the recent dimming event, five more dimming episodes were identified by \cite{anugu_time-evolution_2024} based on visual observational data from the American Association of Variable Star Observers\footnote{\url{https://www.aavso.org/}} (AAVSO) and Digital Access to Smithsonian Collections History (DASCH) archives, during which the visual magnitude of the star decreased by up to 1.2~mag below normal. The average timespan of such a dimming episode has been around three years.

The Great Dimming of RW\,Cep was reported in December 2022 by \citeauthor{ATel15800}, when the star's V magnitude had decreased from $\sim6.4$~mag down to 7.6~mag. In the past 100 years, this recent event seems to be the most significant dimming episode, so calling it the “Great Dimming” seems very appropriate \citep{anugu_time-evolution_2024}. On several occasions \citep[e.g.][]{ATel15800,Leadbeater2023,Anugu2023,anugu_time-evolution_2024}, there have been drawn parallels between the dimming of RW\,Cep and the dimming event of Betelgeuse in 2019 \citep{montarges_dusty_2021}. 

\cite{Leadbeater2023} was among the first to publish spectroscopic monitoring observations taken during the dimming minimum from December 2022 to February 2023, when the star's brightness was at its lowest. Spectra taken over these months reveal new emission components superimposed on the absorption lines -- especially a strong emission in H$\alpha$. The star's spectral class during the dimming minimum was estimated to be K4~I. 

Interferometric imaging of the surface of RW\,Cep in \textit{H} and \textit{K} filters with the Center for High Angular Resolution Astronomy (CHARA) array during and following the Great Dimming reveals an asymmetric stellar surface, extended plume emissions and localised emission and absorption due to dust. The irregular shape of the star may be caused by asymmetric and locally variable mass-loss rate \citep{Anugu2023}. The variable brightness of the stellar surface following the Great Dimming is interpreted as giant convection cells \citep{anugu_time-evolution_2024}. Indeed, \cite{josselin_atmospheric_2007} elaborated on the dimensions and velocity variations in giant convection cells in red supergiants (RSGs). They found that their measurements supported the idea that there are at most a dozen such cells on the surface of a RSG, and changes in these cells would cause irregular light variations in the star. \cite{anugu_time-evolution_2024} estimated the stellar radius in the near-infrared (NIR) at 1100~$R_\odot$.

From NIR spectra in the \textit{K} band \cite{Anugu2023} found that compared to the pre-dimming spectra, there is added flux in the K-band that corresponds to dust emission. Additionally, the \ion{CO} 2.29 $\mathrm{\mu}$m line appears significantly deeper, indicating a lower photospheric temperature. They estimated the temperature from the pre-dimming spectrum taken in 2005 to be 4200~K, and during the dimming 3900~K or even less. 

\cite{humphreys_episodic_2022} explained the cause for the major dimming events in RSGs with magnetic fields combined with surface activity releasing matter from the stellar surface. \cite{anugu_time-evolution_2024} suggested the same mechanism for RW\,Cep. As the released gas travels outward, it condenses into dust, which results in a decrease in the star's brightness. When the cloud dissipates or moves away from the front of the star, the brightness returns to normal levels.

In Sect.~\ref{sect:observations} we give an overview of our observational data, in Sect.~\ref{sect:light_curves} we summarise the photometric variability of RW\,Cep during the Great Dimming to provide the background for our spectroscopic study. In Sect.~\ref{sect:dynamics} we will dive deeper into the radial velocity (\rv) variability at different depths of the vast atmosphere of RW\,Cep and in Sect.~\ref{sect:spec_line_profiles} we will look at specific examples of spectral lines and how the extraordinary circumstances of the Great Dimming have affected them.

\section{Observations}\label{sect:observations}
\subsection{Spectroscopy}

We have collected spectroscopic data from five instruments, covering the years 1999--2024, though there are very few observations from 2007-2021. In this section, we give a brief overview of all spectroscopic observations used in this paper, the individual observations are listed in Appendix~\ref{app:observations}. 

We are using archival spectra from the beginning of the 2000s observed with ELODIE echelle spectrograph, which was operated at the 1.93~m telescope in Observatoire de Haute-Provence \citep{moultaka_elodie-sophie_2004}. The spectra cover the wavelength range 3850--6800~\AA, with a spectral resolution of R$\sim$42\,000 \citep{baranne_elodie_1996}. The reduced spectra were accessed through the ELODIE archive\footnote{\url{http://atlas.obs-hp.fr/elodie/}}. The signal-to-noise ratio (S/N) of the spectra is 200--300.

Additionally, we include observations taken with the Dominion Astrophysical Observatory (DAO) McKellar spectrograph at the 1.2~m telescope \citep{monin_robotic_2014}. The data were accessed through the Canadian Astronomy Data Centre archive\footnote{\url{https://www.cadc-ccda.hia-iha.nrc-cnrc.gc.ca/}}. The spectra have resolution R$\sim$17\,000, the wavelength ranges of spectra vary slightly, but generally cover the range 6300--6900~\AA. The S/N of the spectra is 100--300. 

Two high resolution (R$\sim$80\,000) and very high S/N (>600) spectra from 2006 are from the Polarbase\footnote{\url{http://polarbase.irap.omp.eu/}} archive \citep{petit_polarbase_2014,donati_spectropolarimetric_1997}. Observations were done with the Echelle SpectroPolarimetric Device for the Observation of Stars (ESPaDOnS) and cover a very wide wavelength range from 3700 to 10\,500~\AA. 

We included one observation epoch from the 2-meter telescope in Ond\v{r}ejov Observatory in the Czech Republic, operated by the Astronomical Institute of the Czech Academy of Sciences. The spectra were taken in wavelength range 6250--6750~\AA, with resolution R$\sim$40\,000 and S/N 330. 

We observed RW\,Cep during its dimming period with the high-resolution FIbre-fed Echelle Spectrograph (FIES) \citep{telting_fies_2014} at the Nordic Optical Telescope on La Palma. The spectra were taken in medium-resolution (R$\sim$45\,000) mode covering the wavelength range 3630–8980~\AA. The spectra were reduced with the standard FIEStool pipeline \citep{stempels_fiestool_2017}. The S/N is 200--300.

The bulk of the spectra used in this paper are from Tartu Observatory (TO). Observations were made with the long-slit spectrograph ASP-32 mounted in the Cassegrain focus of the 1.5~m telescope AZT-12. The monitoring program began in December 2022, at the brightness minimum of RW\,Cep, and the latest spectra included in this paper were taken 1.5 years later, in May 2024. The spectra were taken using the 1800~lines\,mm$^{-1}$ grating with the spectral resolution R$\sim$10\,000, covering the wavelength range 6300--6600~\AA{} \citep{folsom_rare_2022}. The S/N of spectra is 200--300. The data were reduced using the \textsc{IRAF}\footnote{\url{https://iraf.net/}}\citep{tody_iraf_1986} packages: \textsc{noao}, \textsc{imred}, \textsc{ccdred}, \textsc{ctioslit} and \textsc{rv}. Wavelength calibration was made based on the ThAr hollow cathode lamp. The spectrograph is mounted at the Cassegrain focus of the telescope, which causes minor flexure in the instrument due to telescope movement. This affects the accuracy of the wavelength scale and the resulting error is wavelength dependent, larger ($\sim$0.1\,\AA) near 6300~\AA, and near-zero at 6600~\AA. For radial velocity measurements of a single line, this can be easily remedied by selecting nearby lines with known wavelengths and measuring the difference (DIB at $\lambda6379.01$ was used in our previous work \citep{kasikov_yellow_2024}). However, in this paper, we measure the radial velocities of a substantial amount of lines over the entire spectral range, so this single-line approach is inadequate. 

To remedy the instrumental wavelength error, we calculated a correction curve. We measured the wavelengths of five relatively strong night sky emission lines in the sky background of a spectrophotometric standard star \object{10 Lac} and compared them to the wavelengths from the European Southern Observatory UV-visual echelle spectrograph (UVES) \citep{hanuschik_flux-calibrated_2003}. Based on the wavelength differences, we calculated a linear correction curve. To correct the spectra of RW\,Cep, we scaled the correction curve to each individual spectrum using the strongest atmospheric [\ion{O}{I}] $\lambda6363.776$ line. Measuring all five lines for each spectrum of RW\,Cep is not feasible, because the other four are much weaker and only become visible in long (>1 h) exposures. Following this wavelength scale correction, the radial velocities measured from TO spectra agree very well to those measured from FIES spectra at nearby dates. We can guarantee velocity accuracy of approximately $\sim2.5$~\kms. 

Spectra were normalised to (pseudo-)continuum with the \textsc{SUPPNet} software \citep{rozanski_suppnet_2022} using a series of carefully selected "anchor points" in the spectra. 

\subsection{Photometry}

We have collected photometric data from several freely accessible archives, given below. A brief introduction to these datasets and the light curves themselves will be discussed in Sect. \ref{sect:light_curves}. 

Kamogata/Kiso/Kyoto Wide Field Survey (KWS) \citep{maehara_automated_2014} observations were taken from 2011 to 2024. The data includes measurements in Johnson-Cousins \textit{V} and in Cousins \textit{Ic} filters\footnote{\url{http://kws.cetus-net.org/~maehara/VSdata.py}}. A total of 416 nights of \textit{V} filter observations and 312 nights of \textit{Ic} filter observations. The mean given instrumental uncertainty of the \textit{V}-filter measurements is 0.006~mag, and for \textit{Ic}-filter observations 0.009~mag. The mean scatter of measurements taken within one night is around 0.011 mag. 

W. Vollmann, a member of the AAVSO (observer code VOL) has observed RW\,Cep using a digital camera in the Johnson-Cousins \textit{V} filter (transformed into standard magnitudes) in 2018--2024 \citep{vollmann_observations_2024}. The mean uncertainty of the observations is 0.013~mag. 

Additionally, we have included multi-epoch photometry from Gaia in \textit{G}, $G_{\mathrm{BP}}$ and $G_{\mathrm{RP}}$ filters \citep{gaia_collaboration_gaia_2016,gaia_collaboration_gaia_2023}. The  uncertainties given for \textit{G} band are $\sim 0.3$~mmag, for $G_{\mathrm{BP}}$-band $\sim 0.9$~mmag and the $G_{\mathrm{RP}}$-band $\sim 0.6$~mmag.

\section{Light curves}\label{sect:light_curves}
\subsection{Data preparation}
The KWS data were cleaned by removing points with instrumental uncertainties larger than 0.03~mag and by filtering the data with a 10-day running median window and removing points that deviated by more than 3$\sigma$. Additionally, observations taken within one night were averaged and nights, where the scatter of measurements in \textit{V} filter was >0.04~mag or \textit{Ic} filter >0.05~mag were excluded. These values are significantly larger than the mean scatter within one night, which is around 0.011~mag for both filters. We also adopted this value as uncertainty displayed in the figures for KWS, as it better describes the scatter of data points than the instrumental uncertainty given for each individual point. The mean uncertainty for (\textit{V-Ic}) colour is 0.017~mag.

The magnitudes in the AAVSO database were systematically slightly lower than those reported by KWS -- a mean difference of $\sim$0.08\,mag. To line up the data from AAVSO and KWS, we shifted all measurements from AAVSO by this value.

\subsection{Results}

\begin{figure}
    \centering
    \includegraphics[width=\hsize]{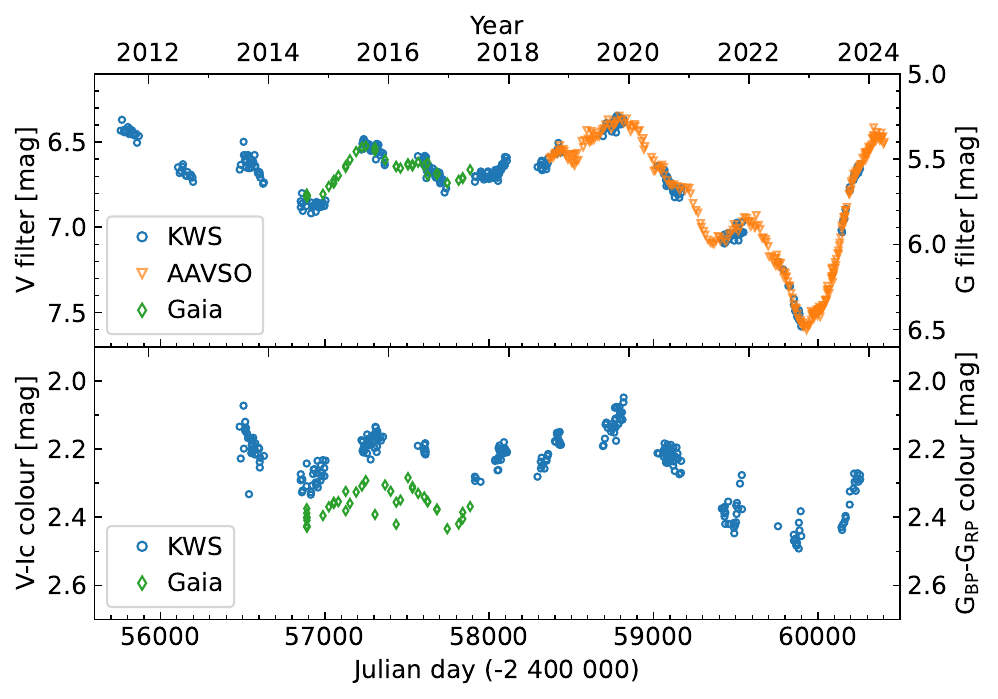}
      \caption{Upper panel: \textit{V}-filter light curve from AAVSO and KWS data, including \textit{G}-filter data from Gaia. Lower panel: (\textit{V-Ic}) colour together with (\textit{$G_{\mathrm{BP}}$-$G_{\mathrm{RP}}$}) colour from Gaia. Each point corresponds to one night. The uncertainties are smaller than the data points. Please note that the y-axes for Gaia data are on the right of the figure, but the scale unit sizes are the same for both \textit{V}-filter and Gaia data. However, the axis limits for \textit{V} filter and Gaia are different in the upper panel.}
    \label{Fig:VIlightcurve}
\end{figure}

The \textit{V}-filter data from AAVSO and KWS have been published and discussed in several previous papers about RW\,Cep \citep{anugu_time-evolution_2024,Anugu2023,ATel15800,Leadbeater2023}. On Fig.~\ref{Fig:VIlightcurve} upper panel we have plotted the AAVSO light curve by W. Vollmann, the smoothed KWS \textit{V}-filter data and the Gaia G-band magnitude. The Gaia data have been plotted on a separate y-axis on the right, the increments of both of the y-axes are the same. Gaia data and AAVSO measurements show smooth variability of brightness. In 2012--2020, the brightness of RW\,Cep in \textit{V} filter was variable between 6.4--6.9 mag. The Great Dimming of RW\,Cep lasted from 2020--2024. The dimming minimum was from Dec 2022 to Jan 2023, during which the brightness dropped to around 7.6 mag. The recovery from the dimming minimum was smooth, without significant bumps or pauses. By mid-2024 the star's brightness was back to normal pre-dimming levels. Our temporal coverage of spectroscopic observations begins near the time of dimming minimum and covers the entire period of brightness recovery.

\begin{figure}
    \centering
    \includegraphics[width=\hsize]{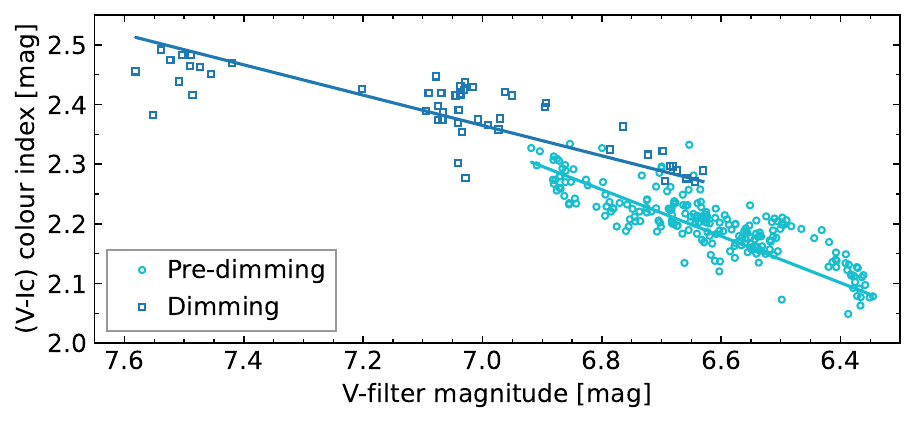}
    \caption{Correlation between \textit{V}-filter magnitude and (\textit{V-Ic}) colour index based on KWS data and two linear fits: one corresponding to normal variability and the other to the dimming period data. The standard deviation of the fit residuals is 0.03\,mag for the pre-dimming period and 0.04\,mag for the dimming period. For (\textit{V-Ic}) colour, KWS has an instrumental uncertainty of about 0.025 mag.  } 
    \label{fig:V-Icorr}
\end{figure}

In the bottom panel of Fig.~\ref{Fig:VIlightcurve}, we have plotted the (\textit{V-Ic}) colour index from KWS and (\textit{$G_{\mathrm{BP}}$-$G_{\mathrm{RP}}$}) colour from Gaia. Both colour indices show similar variability behaviour during the pre-dimming period, between 2.1--2.4~mag for (\textit{V-Ic}) colour and 2.3--2.45~mag for (\textit{$G_{\mathrm{BP}}$-$G_{\mathrm{RP}}$}) colour. During the dimming, the star reddened by about 0.2--0.3~mag based on the KWS (\textit{V-Ic}) colour measurements. The change in colour is accompanied by a change in the effective temperature of the star, as found by \cite{Anugu2023}. There is a very good correlation between the brightness change in \textit{V} filter and the colour index, as seen in Fig. \ref{fig:V-Icorr}. The slope of the correlation changes during the dimming, which is another indicator of the changed state of RW\,Cep.

There have been some estimates of the periodicity in the light curve of RW Cep. \cite{anugu_time-evolution_2024} found a period variable between 350--800 days, with a peak at 403 days based on AAVSO data, and also a longer $\sim$6.8-year period. Similar periods have been described previously in the literature. \cite{percy_studies_2000} used data from the Hipparcos satellite and found a period of 600 days. \cite{kholopov_combined_1998} gave a period of 346 days with visual brightness varying between 8.6 and 10.7~mag. \cite{chinarova_catalogue_2000} found values between 600--900 days and some longer periods exceeding 2000 days for data between the years 1936--2000. During that time the visual brightness of RW Cep varied between 6.4--8.0~mag based on data from Variable Star Observers League in Japan (VSOLJ) and The French Association of Variable Star Observers (AFOEV).

\section{Atmospheric dynamics}\label{sect:dynamics}

We explore the changes in the atmosphere of RW\,Cep during the Great Dimming event based on radial velocity analysis of selected well-measurable spectral lines. We follow the star's variability thanks to the high temporal cadence time series starting from the dimming minimum and continuing up to 1.5 years post-minimum. Additionally, we place the Great Dimming in a historical context, by comparing this event to the behaviour of RW\,Cep in the beginning of 2000s.  We will have a closer look into the variability of H$\alpha$ to investigate the behaviour of the stellar atmosphere during the Great Dimming. The backbone of the analysis is based on the \ion{Fe}{I} lines, which are the most numerous lines in the 6300--6600~\AA{} wavelength region of TO spectra, and we strive to cover a wide range of lines with different intensities and energies of the lower level (\Elow). In addition, we will have a more detailed look at a \ion{Si}{II} line, which is usually found in stars much hotter than RW\,Cep.

\subsection{Methods}

Our best temporally sampled time series begins at the dimming minimum, the observations were made with TO long-slit spectrograph in the wavelength range of 6300--6600~\AA. Therefore, most of the spectral lines discussed in this paper will fall in that region.

We measured the radial velocities of spectral lines by using Gaussian fitting of the line profile core or by calculating the first moment (intensity weighted velocity) of the spectral line. The Gaussian method was preferred for TO, DAO and Ond\v{r}ejov spectra, where the spectral line profiles were more symmetric and consequently the central parts of the line profiles could be fitted well with a Gaussian shape. 

The moment method was the preferable method used for FIES, ELODIE and ESPaDOnS spectra, where the higher spectral resolution reveals the finer structure and asymmetry of the spectral lines. Measurements of the first moment can be significantly affected by any blends in the line wings. The spectrum of RW\,Cep is densely packed with spectral lines, therefore in many cases deblending was necessary before we could measure the line moment. To deblend two lines, we fitted a Gaussian to the line next to the target line and subtracted the fit. When deblending was not possible, but the centre of the line was symmetric enough to be approximated with a Gaussian, we occasionally opted for Gaussian fitting in the case of FIES and ELODIE spectra as well.
In the case of a strong blend and asymmetric spectral line, the radial velocity could not be reliably measured. 

To quantify the uncertainty of moment measurements, we identified that the most significant contribution to the error of the first moment is the selection of the wavelength window in which it is calculated. Selecting the appropriate window for each individual spectral line is critical to getting accurate results. We estimated the error statistically. We measured the line moment in a window that extended past line wings or included only the central part of the line. In all ELODIE, ESPaDOnS and FIES spectra, we measured several relatively unblended spectral lines (e.g. \ion{Si}{II} $\lambda$6347 and \ion{Fe}{I} $\lambda$6358) using around ten different window widths. The standard deviation of moments calculated in different wavelength windows was $\sim1.2$ \kms, though in cases where the S/N of the spectrum is low ($\sim100$), or the line is weak (<20\,\% continuum depth), the error can be up to $\sim2$ \kms. 

Additionally, we measured the depth of each spectral line. We define the depth of spectral line in respect to the continuum, and higher values correspond to stronger absorption lines. Line depth is a more robust measure of the intensity of a spectral line than the equivalent width. Equivalent width can be complicated to measure when the lines are blended -- as is often the case in RW\,Cep -- and if the continuum is difficult to place. Line depth is also affected by inaccuracies due to the position of the continuum, but we have made all efforts to bring it to a minimum by carefully selecting the points through which the pseudo-continuum was placed for the spectra of RW\,Cep. We estimate the precision of intensity for line depth measurements at 0.006--0.008. We measured the line depth either through Gaussian fitting in the cases where it was used for radial velocity or by measuring the lowest point in the spectral line. An important note is that, unlike equivalent width, line depth is sensitive to the resolution of the instrument. This makes it more difficult to compare results from different instruments. In the case of RW\,Cep, this is not a significant issue due to its extremely wide spectral lines, so even with moderate resolution, the line profile is physical rather than instrumental. 

\subsection{Following H$\alpha$ through the dimming}

\begin{figure*}
\sidecaption
   \includegraphics[width=12cm]{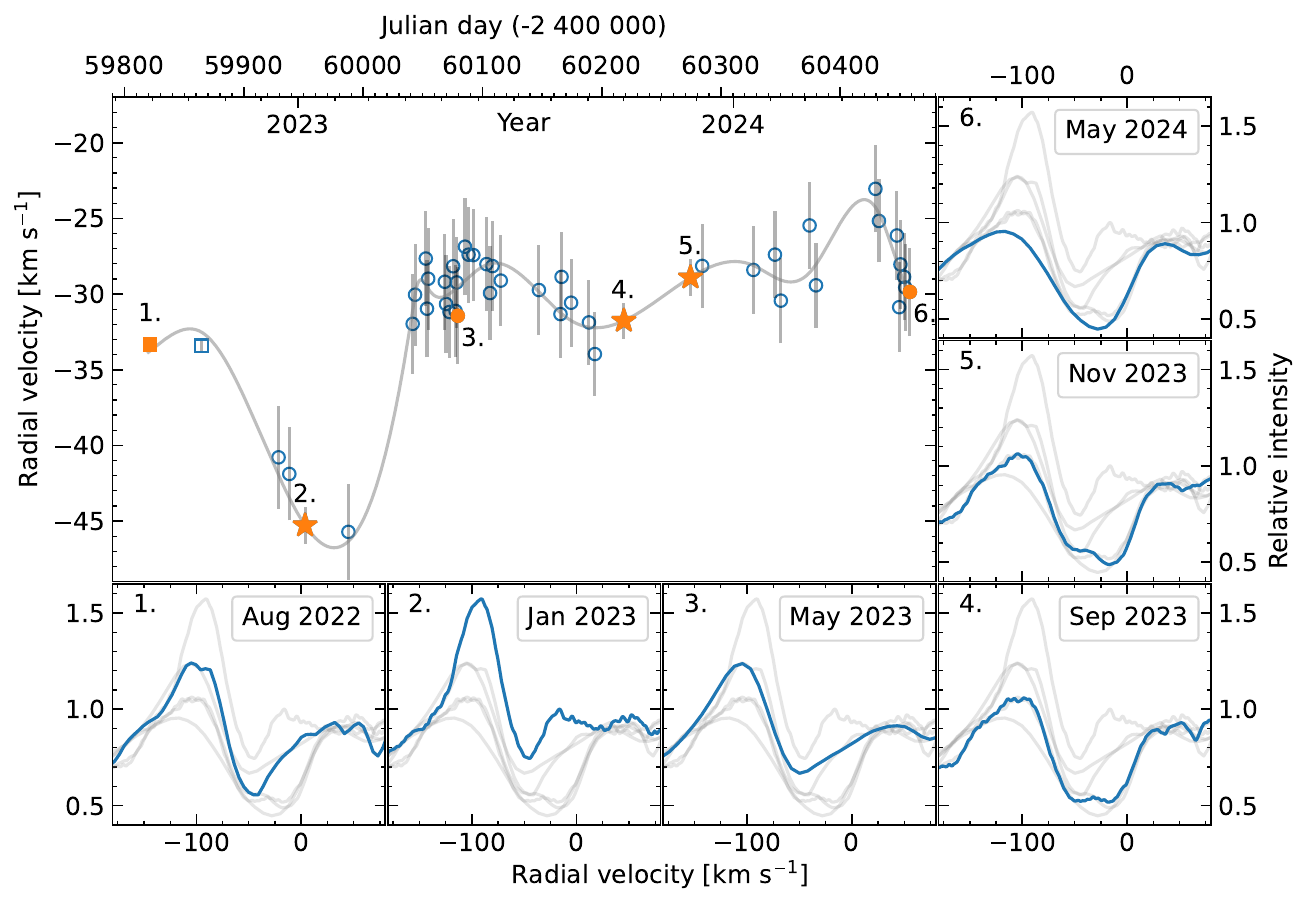}
     \caption{Variability of the radial velocity and H$\alpha$ line profile during the Great Dimming. The main panel showcases the radial velocity variability, and the small numbered panels around it correspond to spectral line profiles on dates highlighted and numbered on the radial velocity curve of the main panel. Behind the highlighted line, all other five profiles are drawn in light grey, to provide a point of reference.   Panel nr. 1 is from before the dimming minimum, but the emission is already visible. Panel nr. 2 showcases the profile when the star was faintest. The following panels 3--6 are taken when the star was already brightening, the strength of the emission in H$\alpha$ decreased and the variability in radial velocity is caused mostly due to the variable absorption component. Panels 2, 4 and 5 are FIES spectra, panel 1 is from DAO and panels 3 and 6 are TO spectra, the lower resolution of the latter two results in a smoother profile shape. Please note that the systemic velocity of the star has not been subtracted from the velocity axis.}
     \label{Fig:Halpha}
\end{figure*}

Figure~\ref{Fig:Halpha} gives an overview of the variability of the radial velocity (first moment) of the H$\alpha$ absorption component during and after the dimming minimum. Some example line profiles are given on the small panels at the sides of the main panel. The H$\alpha$ profile of RW\,Cep is non-symmetric and slightly variable in the 'calm' state of the star based on spectra from 1999--2016. Its radial velocity during that time varied between $-55$ to $-40$~\kms. An extended timeline of the variability of the H$\alpha$ profile in RW\,Cep is given in Appendix~\ref{app:halpha}. In the appendix, we show line profiles from the beginning of the 2000s, as well as a more densely temporally sampled series of changes in H$\alpha$ during the Great Dimming. 

The Great Dimming of 2022--2023 introduced a strong emission component to the line profile of H$\alpha$ that spans a wide range of velocities. There was already a hint of emission more than a year before the dimming minimum -- in a spectrum from September 2021 (panel 7 in App. \ref{app:halpha}, compare the blue wing with previous epochs), when the blue wing of the H$\alpha$ line reached almost 10\% above the continuum level. \cite{anugu_time-evolution_2024} suggested, based on light curve analysis, that the gas ejection began in 2020. In September 2021, the brightness of RW\,Cep was already rapidly decreasing (see the light curve in Fig.\,\ref{Fig:VIlightcurve}). However, at that time, the radial velocity of H$\alpha$ was at a normal level ($-42$~\kms). The brightness decrease indicates that around August 2021, the ejected gas was already condensing into dust, which came into our line of sight and began shielding starlight. The movement of the cloud was perhaps due to the rotation of the star, simply a physical movement, or a combination of both. The appearance of the cloud of gas and dust gave rise to additional blue-shifted emission components that we see in many spectral lines. The main panel of Fig. \ref{Fig:Halpha} starts about a year later, from August 2022, when the emission component in the blue wing of H$\alpha$ was readily visible (panel no. 1).  

The emission strength increased in the following months while the star's brightness decreased. The emission reached its maximum in the beginning of January 2023 (panel no. 2), during the brightness minimum. The peak of the emission was located blueward (at lower velocities) of the central absorption, resulting in the asymmetric profile. The emission width extended from near {$-120$~\kms} to velocities up to 0~\kms. In the figure, the wings of the emission component can be seen on both sides of the photospheric absorption. 

By May 2023 (panel no. 3), the emission strength decreased, and the red wing of the absorption line became very broad, resulting in an increase in the measured radial velocity. In the following panels 4 and 5, we see a further decrease in the strength of the emission component and a strongly asymmetric tilt of the absorption line towards higher velocities. The high-resolution FIES spectra shown in panels 4 and 5 reveal the intricacies in the bottom of the absorption line, the spectrum on panel 6 has a smoother profile shape due to lower resolution. From interferometric imaging of \cite{Anugu2023}, it was visible that by October 2023 the shape of the stellar surface had become much more symmetric. Based on the spectral features of H$\alpha$, the dust/gas ejected during the eruption had not yet fully dissipated.

The emission in H$\alpha$ was down to continuum level by the beginning of 2024 (panel no. 6), when the star regained its former brightness. However, in mid-2024 the H$\alpha$ line profile has not yet recovered its former depth or pre-outburst radial velocity. This could still be the influence of the weakening emission. 

In Betelgeuse, the surface mass ejection was triggered by outward-directed photospheric motion and shocks \citep{dupree_great_2022}. However, for RW\,Cep in Sept 2021, the measured {\rv} values of H$\alpha$ and other lines (discussed in the following sections) remain within the normal range, so if any outburst activity took place in 2021, then either it did not significantly disrupt the stellar atmosphere, or it was localised to the other side of the star and thus not visible in our {\rv} measurements. Within the next year, the gas and dust cloud came further into our line of sight and, by the time of the brightness minimum, shielded the largest percentage of the stellar disc \citep[as can be seen in interferometric images by][]{Anugu2023} and the emission in H$\alpha$ reached its maximum strength. Following the dimming minimum, the cloud either dispersed or moved out of the line of sight, resulting in a smooth increase of the star's brightness and accordingly, a decrease in H$\alpha$ emission strength.

\subsection{\ion{Fe}{I} radial velocity results}

\begin{figure*}
    \centering
    \includegraphics[width=17cm]{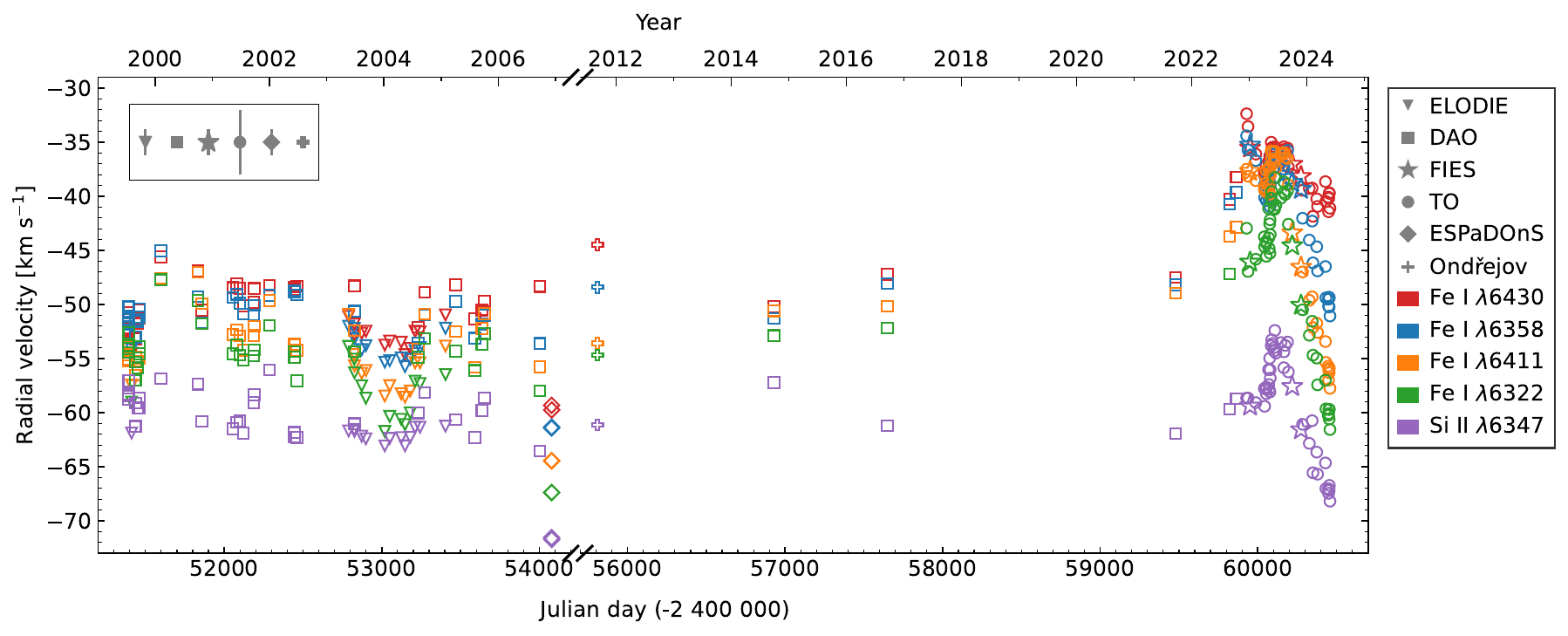}
    \caption{Mean core velocities of four representative \ion{Fe}{I} lines and a \ion{Si}{II} line. The full timeline covers the years from the beginning of the 2000s until mid-2024. Please note the gap in the x-axis. Different symbols mark data from different spectrographs (mean error bars for each instrument are given in the top-left corner). The colours red, blue, orange, and green correspond to \ion{Fe}{I} $\lambda\lambda$6430,6358,6411, and 6322 lines respectively, and purple corresponds to \ion{Si}{II} $\lambda$6347. A detailed look at each individual \ion{Fe}{I} line during the dimming event is given in the Appendix~\ref{fig:FeIradveldetailed}.}
    \label{fig:FeIradvel}
\end{figure*}

We measured radial velocities of 33 selected \ion{Fe}{I} lines, the details of these lines are given in Table~\ref{table:FeIlines}. The lines without an asterisk are well-resolved in TO spectra (16 lines in total) and we followed their variability through the dimming event. The lines marked with an asterisk in the table either are not within TO spectral region or are not well-resolved, but they are included in the comparative analysis of the historic radial velocity behaviour in Sect.~\ref{sect:radvel_hist_comp}.

We selected relatively unblended lines, allowing for reliable measurements of their radial velocities and line depths. The spectral line data for all selected lines (wavelength, excitation potential of the lower level {\Elow} and oscillator strength $\log gf$) are from the Belgian Repository of fundamental Atomic Data and Stellar Spectra (BRASS) database\footnote{\url{http://brass.sdf.org/}} \citep{lobel_belgian_2019}. 

In Fig.~\ref{fig:FeIradvel}, we have drawn a full timeline of the radial velocity variability of four distinct \ion{Fe}{I} lines $\lambda\lambda$6430,6358,6411, and 6322 from 1999 to 2024. These four lines are representative subset of our sample, their {\Elow} values vary from 0.8--3.6~eV. Additionally, we have added the \ion{Si}{II} $\lambda$6347.11 line to the figure, which is an interesting line in many ways and its radial velocity is much lower than that of the \ion{Fe}{I} lines. The {\Elow} of \ion{Si}{II} is 8.121\,eV and $\log gf$ is 0.297. We include some additional comments on the \ion{Si}{II} line in Sect.~\ref{Sect:comments_on_SiII}.

At the beginning of 2000s, the spectral lines span a range of velocities from $-63$ to $-45$~\kms, the variability amplitudes of individual lines are in {5--10~\kms} range. ELODIE data in 2003--2005 reveals a small dip in the radial velocity curve. However, comparison with the long-time visual brightness curve given by \cite{Anugu2023} shows, that the brightness of RW\,Cep was very stable at that time between 6.8--7~mag. At the same time, the radial velocity of \ion{Si}{II} shows barely any change. There are a few more minor jumps in radial velocity in the beginning of 2000, we will return to this later in Sect.~\ref{sect:radvel_hist_comp}. 

In the intermediate period from 2006 to 2021, we have very little data. Please mind the gap in Fig.~\ref{fig:FeIradvel} between the years 2007--2011. The Ond\v{r}ejov spectrum from 2011 and DAO spectra from 2014, 2016 and 2021 show that the radial velocity remained relatively on the same level as before. 

Throughout the years, we see a clear gradient in the radial velocities of different spectral lines. Some lines, such as \ion{Si}{II} and \ion{Fe}{I}~$\lambda$6322 have systematically lower radial velocities than other lines, e.g. \ion{Fe}{I}~$\lambda6430$.
Even though the radial velocities of individual lines are variable over time, all lines vary almost in synchronisation with each other during the velocity drops and rises that we see over the years (see Sect.~\ref{sect:sysvel} and \ref{sect:radvel_gradient_discussion}).

\begin{figure}[h!]
    \centering
    \includegraphics[width=\linewidth]{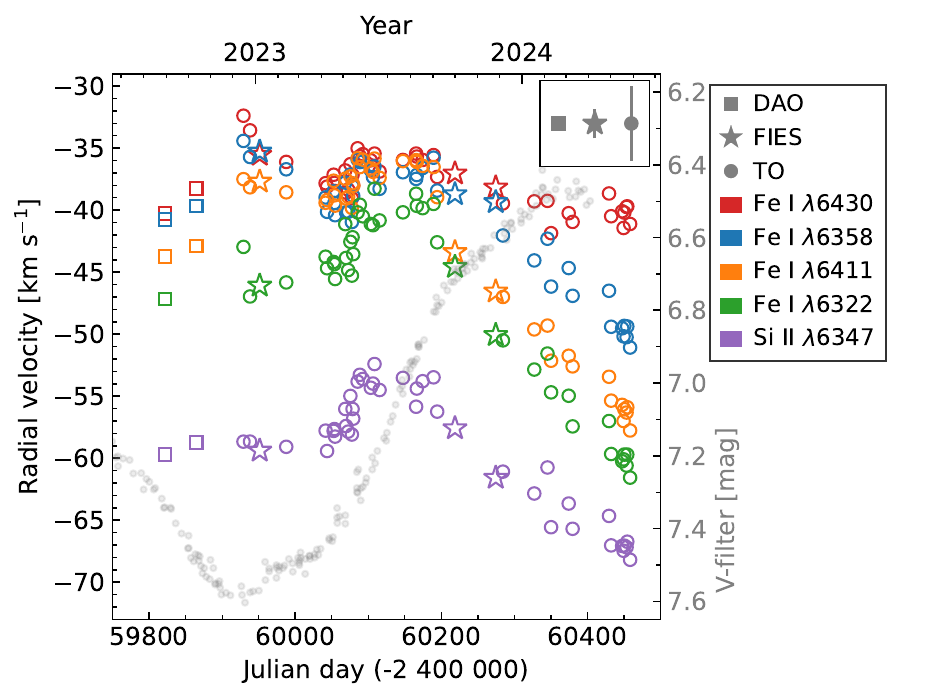}
    \caption{Zoom in on the dimming period from Fig. \ref{fig:FeIradvel} with the same spectral lines using the same labels and colours. Light grey shows the \textit{V}-filter light curve for reference (y-axis on the right). All four \ion{Fe}{I} lines achieve similar maximum radial velocities during the dimming minimum, but the post-dimming behaviour is different: some lines have returned to their normal levels (orange and green in the figure), while others (red) retain the high radial velocity for longer. }
    \label{fig:FeI_radvel_zoom}
\end{figure}

In the context of the full radial velocity curve, the Great Dimming brings a sharp rise in velocity.  A closer look at the curve during Great Dimming is given in Fig.~\ref{fig:FeI_radvel_zoom}. We show the radial velocity curves of four representative \ion{Fe}{I} lines, the \ion{Si}{II} line, along with the brightness in the \textit{V} filter for reference (light grey). For the \ion{Fe}{I} lines, there seem to be two separate velocity maxima -- one corresponds to the dimming minimum in late-2022 and beginning of 2023 (JD $\sim$2\,459\,950) and the other peak approximately six months later, in May-June of 2023 (JD $\sim$2\,460\,100).  For \ion{Si}{II}, there is only one velocity maximum in May-June. During the peak, the velocities of the \ion{Fe}{I} and the \ion{Si}{II} were {10--20~\kms} over the normal level. The initial maximum that corresponds to the dimming minimum is most likely related to the added emission component that we saw in H$\alpha$, but it also affects other lines. The second velocity maximum is half a year later, and the following behaviour of the spectral lines are related to large-scale motions in the atmosphere in RW\,Cep. We show the variability in line profiles at that time in Sect.~\ref{sect:Fe_line_profiles}. 

In the radial velocity curve, we see that after the dimming minimum, the lines have behaved quite differently -- some (such as $\lambda\lambda$6411,6322,6358) have almost returned to their normal velocities within a year after the velocity maximum. However, the velocities of other lines (e.g. $\lambda$6430) still remain almost {10\,\kms} higher than normal. The velocity span of all \ion{Fe}{I} lines post-dimming is $\sim$25\,\kms. 

\subsection{Line depth}

\begin{figure}
    \centering
    \includegraphics[width=\linewidth]{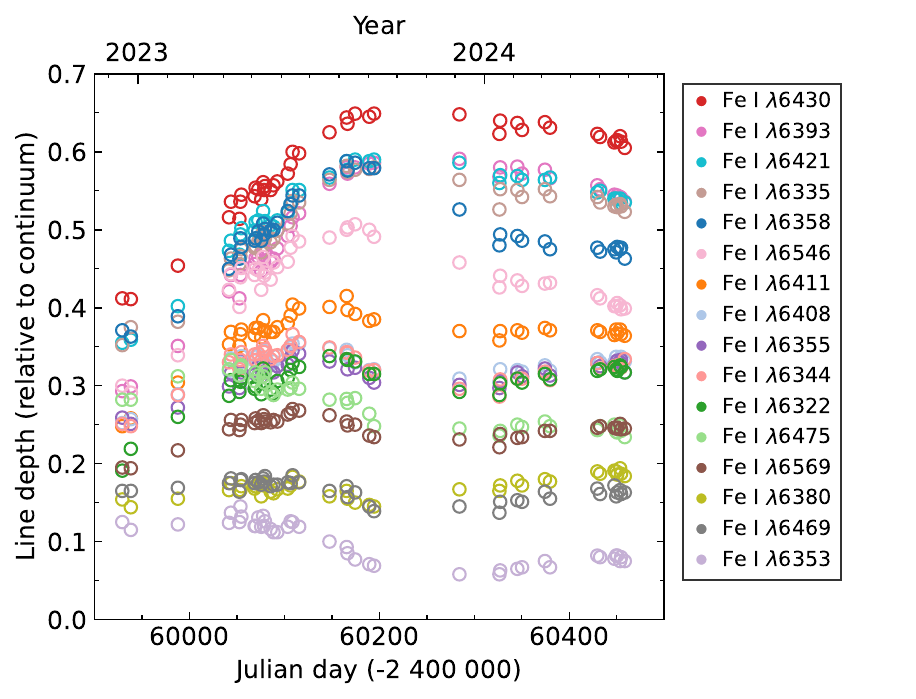}
    \caption{Line depth variability in time measured from TO spectra. Colours mark different \ion{Fe}{I} lines. The \ion{Fe}{I} lines from Fig.~\ref{fig:FeI_radvel_zoom} are marked here with the same red, blue, orange and green colours. Stronger spectral lines have higher line depth values. The error bars are up to the same size as the data points. }
    \label{fig:FeI_line_depth_to}
\end{figure}

We measured the line depths of the selected \ion{Fe}{I} lines from TO spectra during the Great Dimming. Fig.~\ref{fig:FeI_line_depth_to} shows the timeline of variability. Two significant observations can be drawn from this figure: 1) Stronger lines exhibit greater variability in their line depth compared to weaker lines; 2) There is a noticeable shift in the trend of line depth changes in mid- to late-2023, where the increasing line depths appear to plateau, and in the case of weaker lines, even begin to decline. The change in the trend takes place after the radial velocity maximum in mid-2023.

\begin{figure}
    \centering
    \includegraphics[width=\linewidth]{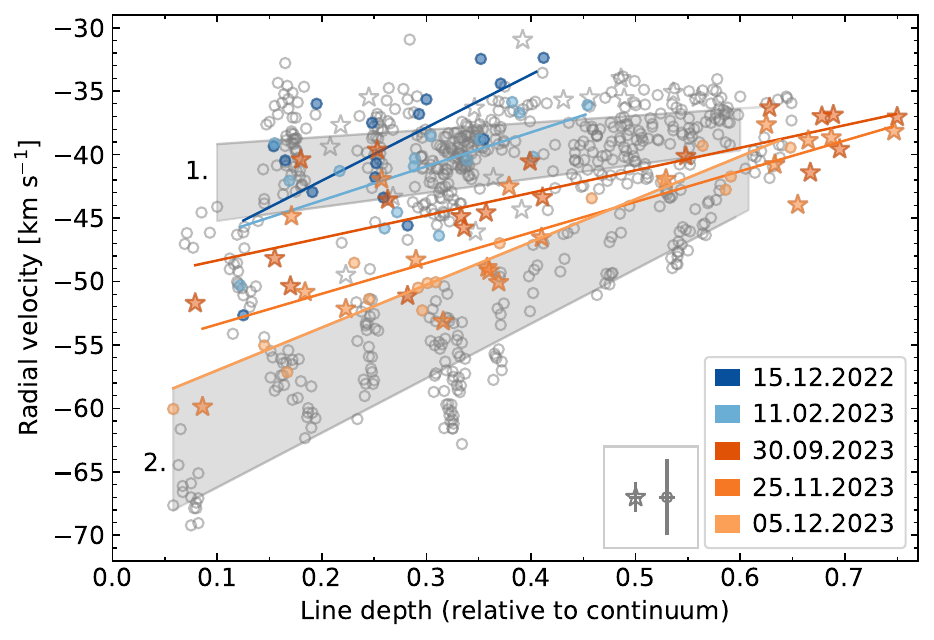}
    \caption{Correlation between line depth and \rv. TO data is marked with circles and FIES data with stars. The grey areas describe the correlation between {\rv} and line depth during two distinct time periods: Area no. 1 is for observations made between March 2023 and Aug 2023; Area no. 2 is for spectra from Dec 2023 onwards. Highlighted dates show different correlation behaviour. The orange-coloured dates are from the time period between those covered by the grey areas: the correlation shift from area 1 to area 2 happened within a few months in late 2023. The blue-coloured dates are from the dimming minimum, when lines were extremely shallow. }
    \label{fig:FeI_line_depth_vs_radvel_to}
\end{figure}

In Fig.~\ref{fig:FeI_line_depth_vs_radvel_to}, we correlate the variability of \ion{Fe}{I} radial velocity with the line depths. We plotted the {\rv} vs line depth on each observation epoch (the grey circles for TO data and stars for FIES data). Next, we calculated a linear fit through the data points for each epoch. The grey areas on the plot mark these linear fits. Area no. 1 describes the relation between {\rv} and line depth for all observations made between March 2023 and Aug 2023; Area no. 2 describes the relation between {\rv} and line depth in observations starting from Dec 2023 onwards. 

We notice a significant change in the slope of the correlation between {\rv} and line depth between the two periods. The correlation shifted in September-November 2023 and we have a couple of FIES observations taken during that time (highlighted in orange on the plot). The line depths of TO and FIES data are not directly comparable due to the different resolutions (spectral lines of TO are shallower and thus FIES data have not been included in Fig.~\ref{fig:FeI_line_depth_to}), but the slope of the correlation is comparable. The change in correlation corresponds to the stabilisation of line depths.  

Additionally, we highlight in blue a couple of epochs from the brightness minimum in late 2022 and early 2023, when the lines were very shallow and the correlation slopes were notably steeper. The Great Dimming significantly affected the properties of the spectral lines, and even though the brightness in the \textit{V} filter has returned to normal levels by mid-2023, the spectrum still reveals the after-effects of this event. 

\subsection{In comparison to the historic data}\label{sect:radvel_hist_comp}

\begin{figure}
    \centering
    \includegraphics[width=\linewidth]{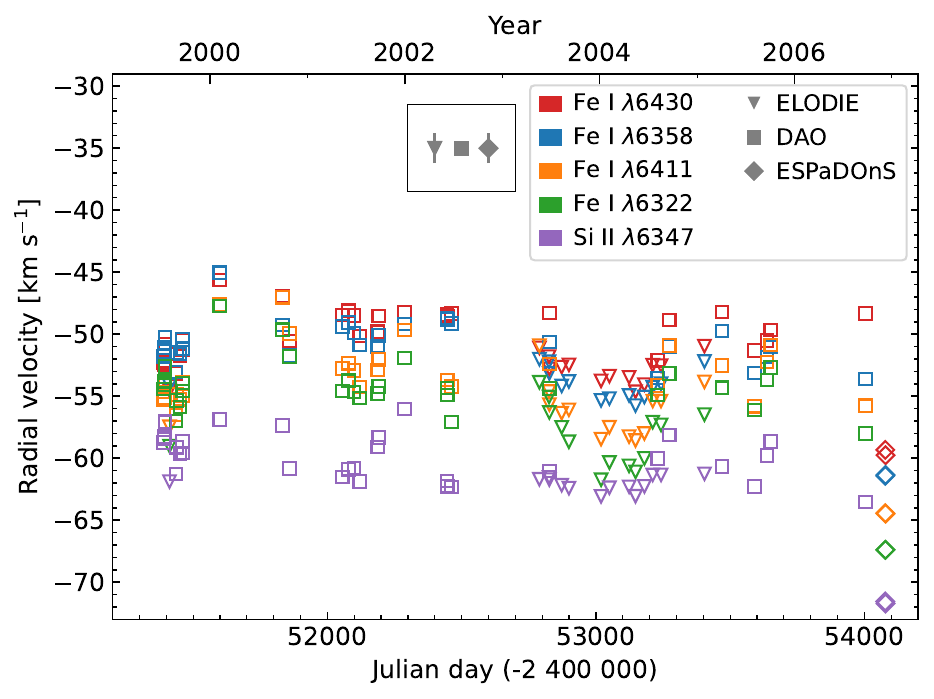}
    \caption{Zoom in on the radial velocity Fig.~\ref{fig:FeIradvel}, focusing on the years 1999--2006. Different instruments are marked with different symbols, and colours correspond to the same spectral lines as in Figs.~\ref{fig:FeIradvel},~\ref{fig:FeI_radvel_zoom}.}
    \label{fig:radvel_early_2000s}
\end{figure}

\begin{figure}
    \centering
    \includegraphics[width=\linewidth]{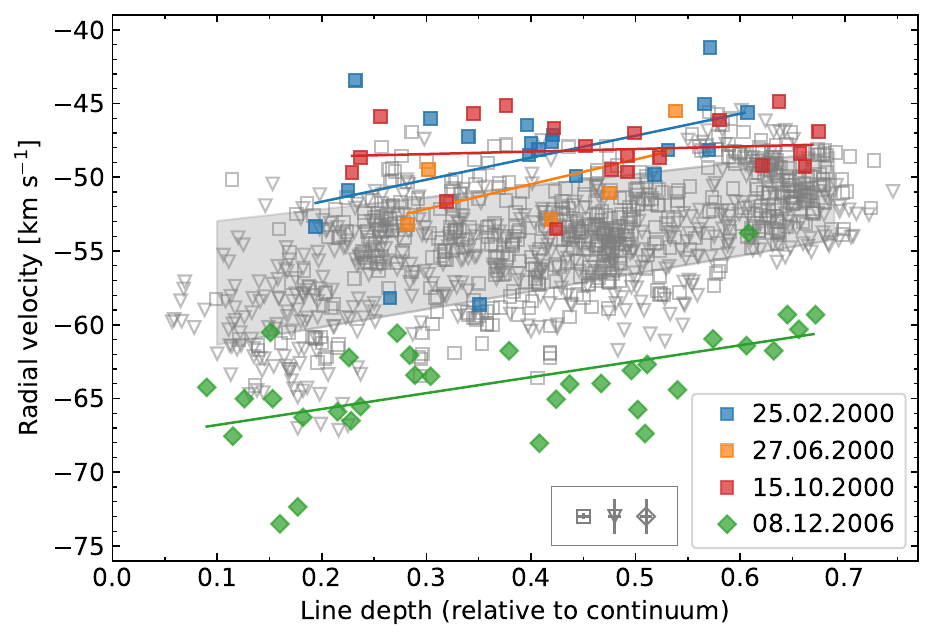}
    \caption{Correlation between {\rv} and line depth in 1999--2006. The grey area shows the standard correlation slopes. Epochs during which the correlation does not fit the standard have been highlighted in different colours. Different shapes correspond to different instruments: square is DAO, triangle is ELODIE and diamond is ESPaDOnS.}
    \label{fig:line_depth_vs_radvel_early_2000s}
\end{figure}

To provide further context to the dimming event, we explore the variability of spectral line properties at the beginning of the 2000s, when the brightness of the star was relatively stable. In Fig.~\ref{fig:radvel_early_2000s}, we have plotted a zoomed-in version of the radial velocity behaviour of RW\,Cep in that period. In this study, we consider this period as the 'normal' or 'calm' state of the star. We can see some minor changes in the radial velocity during that time: the small slope in 2003-2004 that appears from well-sampled ELODIE data; an abrupt {$\sim$10\,\kms} jump in early 2000; and a sudden and significant drop in late 2006, revealed by ESPaDOnS spectra. Similarly to the dimming period, we measured the spectral line depths and drew a correlation plot presented in Fig.~\ref{fig:line_depth_vs_radvel_early_2000s}. 

On most epochs, the correlation between {\rv} and line depth follows the same trend marked with the grey area in Fig.~\ref{fig:line_depth_vs_radvel_early_2000s}. The actual correlation might not be fully linear, especially for the weakest lines, which have a strong preference towards lower velocities. Epochs with distinct correlation slopes have been highlighted with colours. The first group: blue, orange and red correspond to the sudden increase in radial velocity in the year 2000. During that time, the slopes of the correlation are similar to the correlations at the beginning of the Great Dimming. In the light curve \citep[visual observations given in][]{anugu_time-evolution_2024}, we see a minor local minimum at that time with visual brightness 7.3--7.1\,mag. 

The other outlier in the data from the early 2000s are the two ESPaDOnS spectra from December 2006 (one of them is plotted in green on the figure). From these spectra, we measure the lowest {\rv} values for RW\,Cep. Otherwise, the spectra seem similar to others from that period. Regrettably, we do not have any more spectra following the ESPaDOnS measurements. 

\subsection{Determination of the systemic velocity}\label{sect:sysvel}

\cite{gray_observation_2022} has explained that systematic blueward velocity shifts in the stellar atmosphere are the result of granulation: more light originates from the central parts of the granules than the inter-granular areas. Therefore, our radial velocity values measure the granule rise velocity averaged over the stellar disc. With stronger lines, we probe the upper regions of the atmosphere, where the rise velocity nears zero, and with weaker lines, we probe the deeper layers with higher rise velocities. A velocity gradient is entirely expected. The ELODIE data from 2003--2005 were published in \cite{josselin_atmospheric_2007}, where they also wrote that the spectral lines formed in the innermost layers of the atmosphere were systematically blue-shifted compared to the lines formed in the outer layers.

The explanation of the velocity gradient due to granulation (or in our case, giant convective cells) allows us to estimate the systemic velocity ($v_{\mathrm{sys}}$) of RW\,Cep. For the calculation, we use the mean velocities of the strongest \ion{Fe}{I} lines during the relatively calm period at the beginning of the 2000s. There are six lines with line depth values $\geq 0.60$ in Table \ref{table:FeIlines}. We averaged their radial velocities in 1999--2006 (leaving out the December 2006 event) and found an average velocity of $-50.3\pm 3.3$~\kms, which we adopt as the $v_{\mathrm{sys}}$ of the star.

Based on measuring the bisector of about 50 lines in a few-hundred angstrom region near 4200~\AA, \cite{merrill_complex_1956} found an average displacement of $-54$~\kms, which they noted could be the radial velocity of the star. \cite{gahm_spectral_1972} measured the velocity of 40 lines in visual-red and found a mean velocity of $-62.8$~\kms. The different values that these authors have used as the radial velocity of RW\,Cep most likely arise from the selection of lines and different spectral regions used in the analysis. Determining the $v_{\mathrm{sys}}$ of RW\,Cep based on a limited selection of lines in a small number of spectra can lead to these inconsistencies. 

\subsection{Comments on the \ion{Si}{II} line}\label{Sect:comments_on_SiII}

A fascinating property in the spectrum of RW\,Cep is the existence of the \ion{Si}{II} doublet at $\lambda\lambda$ 6347,6371. The \ion{Si}{II} $\lambda6347.11$ in the spectrum of RW\,Cep is relatively weak, and the other line of the same doublet at $\lambda6371.37$ is strongly blended. These lines have rather high {\Elow} values at 8.121\,eV ($\log gf = 0.297$) and are usually found in much hotter stars \citep[examples in YHGs by][]{kourniotis_revisiting_2022,kasikov_yellow_2024}. Therefore, in RW\,Cep, they must form very deep in the vast atmosphere of the hypergiant.  

Different dynamics influencing the \ion{Si}{II} lines are also apparent in their radial velocities, which are {5--10~\kms} lower than those of the iron lines. The radial velocity of \ion{Si}{II} $\lambda6347$ is also less variable compared to the \ion{Fe}{I} lines (See Fig.~\ref{fig:radvel_early_2000s}). During the early 2000s, its radial velocity was near $-60$~\kms, with small fluctuations with an amplitude of a few \kms, the small dip in 2003--2004 that we saw in the \ion{Fe}{I} line is almost not noticeable in the \ion{Si}{II} lines.

During the Great Dimming, the \ion{Si}{II} line does not show a velocity maximum during the dimming minimum (see Fig.~\ref{fig:FeI_radvel_zoom}). That peak appears in stronger \ion{Fe}{I} lines, and we linked it to the added emission component that influences the line profile. The \ion{Si}{II} line shows very little variability during the dimming minimum and no hint of an added emission. Around half a year after the minimum, the \ion{Si}{II} line behaves similarly to the \ion{Fe}{I} lines. \ion{Si}{II} line also reaches a maximum velocity near mid-2023, after which its radial velocity steadily decreases even below its normal value at $-60$~\kms. By May 2024 the {\rv} was down to $-70$~\kms. Therefore, the post-dimming radial velocity behaviour could be linked to dynamical motions that reach into deeper atmospheric layers.

\subsection{Discussion: The velocity gradient}\label{sect:radvel_gradient_discussion}

To describe the atmospheric motions of RW\,Cep before and during the dimming, we relied mainly on the \ion{Fe}{I} lines, which are plentiful in the spectrum of RW\,Cep and in the limited spectral region of TO spectra. The radial velocity time series measurements show the behaviour of the \ion{Fe}{I} lines being different depending on the depth of the line. The weaker \ion{Fe}{I} lines, but also \ion{Si}{II}, have generally high (>2.5\,eV) {\Elow} values and describe the motions deeper in the atmosphere. Stronger lines describe the upper parts of the atmosphere. We have shown a clear correlation between line depth and its radial velocity in Fig.\,\ref{fig:FeI_line_depth_vs_radvel_to} and Fig.\,\ref{fig:line_depth_vs_radvel_early_2000s}. The velocity stratification remains the same during 1999--2006, as well as in the extraordinary conditions during the dimming period. That is clear from Fig.\,\ref{fig:FeIradvel}. However, the long-term stability of the velocities that we measure in RW\,Cep is not always the case for other RSGs. 

\cite{kravchenko_tomography_2019} studied the atmosphere of $\mu$\,Cep, a RSG with a radius of nearly 1000\,$R_\odot$ and effective temperature $T_{\mathrm{eff}}\approx3700$\,K (at first glance a rather similar star to RW\,Cep). They measured the first moments of spectral lines originating at different depths of the star's atmosphere over a time period of almost 7 years. They also detected a velocity gradient in the atmosphere -- lines formed at deeper regions had different velocities than lines formed higher up. However, the velocity gradient they measured was not stable in time: over a timescale of 300--800 days, the gradient reversed. If the upper layers had previously had higher radial velocities, then the velocities decreased and the velocities of deeper layers increased. The star underwent three variability cycles within their observation period. This was considered a clear indication of the movement of giant convective cells and the turn-over of material in the atmosphere. 

RW\,Cep is in both size and temperature very similar to $\mu$\,Cep. Both stars also show long-term periodic variability in their light curves: 800-day and 4400-day periods were cited by \cite{kravchenko_tomography_2019} for $\mu$\,Cep and 400-day and 2500-day periods found by \cite{anugu_time-evolution_2024} for RW\,Cep. Thus, it would be expected that the movement of its giant convective cells \citep[as seen in interferometry by][]{anugu_time-evolution_2024,Anugu2023} would be revealed in its radial velocity field similarly to how we see it in $\mu$\,Cep. However, we do not see such atmospheric turn-over activity in RW\,Cep.

\section{Spectral line profiles}\label{sect:spec_line_profiles}

\cite{merrill_complex_1956} performed a thorough analysis of line profiles seen in the spectrum of RW\,Cep. They noted that the intensities of \ion{Fe}{I} lines belonging to the same multiplets do not show a correlation to the laboratory values. Spectra taken 10 months apart showed only very minor differences. \cite{gahm_spectral_1972} also marked, that independent of their excitation potential, the \ion{Fe}{I} lines are enhanced compared to other supergiants of K spectral class. 

During the dimming minimum, we did not detect any strong TiO bands in the spectrum of RW\,Cep. If present, they are either weak or strongly blended with other lines. The TiO bands were not present in the ELODIE spectra from the beginning of the 2000s either \citep{josselin_atmosphere_2003,josselin_atmospheric_2007}. In contrast to Betelgeuse, where they are always visible \citep{levesque_betelgeuse_2020,harper_photospheric_2020} and the YHG $\rho$\,Cas, where they have been detected during the outburst periods, and they form in the upper atmospheric layers with $T_{\mathrm{eff}}<4000$\,K \citep{lobel_high-resolution_2003}. 

During the dimming, we see added blue-shifted emission over the continuum similar to what we discussed in  H$\alpha$ in many other low-excitation absorption lines: \ion{Ca}{I}, \ion{Ti}{I}, \ion{Sc}{II}, \ion{V}{I}. In other lines, the added emission is not strong enough to cross the continuum, but it nevertheless significantly affects the line shape and its velocity. \ion{Fe}{I} lines are a good example of this. 

\subsection{\ion{Fe}{I} lines throughout the dimming}\label{sect:Fe_line_profiles}

\begin{figure*}
    \sidecaption
    \centering
    \includegraphics[width=12cm]{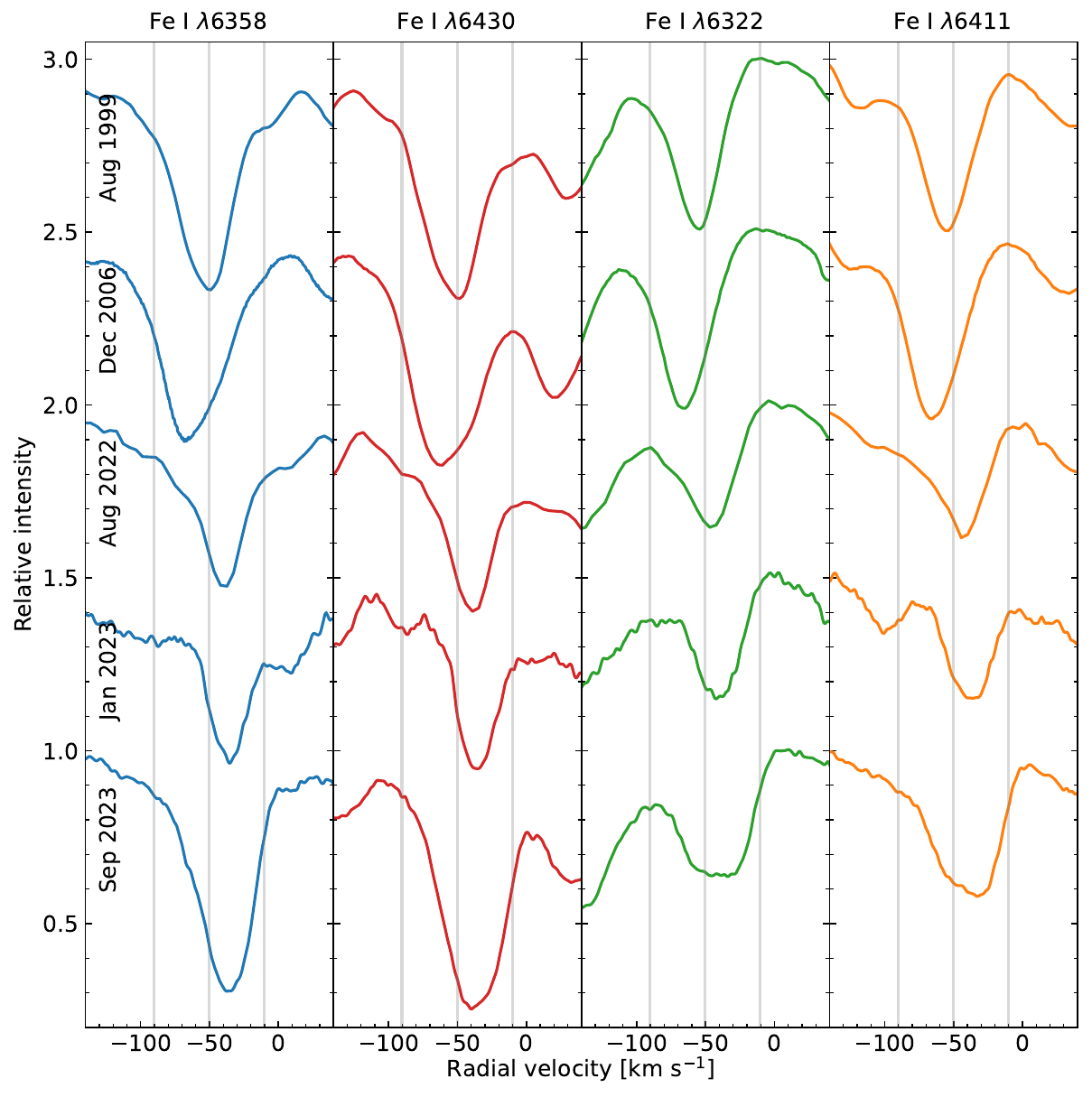}
    \caption{Comparison of four \ion{Fe}{I} lines (columns) on different epochs (rows) before and during the Great Dimming. The $\lambda\lambda6358,6430$ lines represent the strong \ion{Fe}{I} lines and the $\lambda\lambda6322,6411$ represent the weak \ion{Fe}{I} lines. The colours of the lines correspond to the colours in Fig. \ref{fig:FeIradvel}. The central thin grey line traces the $v_{\mathrm{sys}}$ and two additional lines are drawn at {$\pm40$\,\kms} to guide the eye. The spectra are from ELODIE (August 1999), ESPaDOnS (December 2006), DAO (August 2022) and FIES (January and September 2023) The first two spectra from 1999 and 2006 characterise the line properties during the calm state of the star. The August 2022 and January 2023 show the spectra during the Great Dimming and illustrate the effects of the added blue-shifted emission. The added emission shifts the absorption to higher velocities and decreases the line strength and width. The September 2023 spectrum shows the post-dimming behaviour of the lines. The strong lines return to near their normal strength while remaining at higher velocities. The weaker lines gain a significantly wider line profile and remain shallower, but their radial velocity returns to near pre-dimming levels.}
    \label{Fig:FeIlineprofiles}
\end{figure*}

On Fig.~\ref{Fig:FeIlineprofiles}, we have drawn the line profiles of four \ion{Fe}{I} that were also featured on the radial velocity timeline Fig.~\ref{fig:FeIradvel} (the colour palette is the same). Before the dimming, all \ion{Fe}{I} lines displayed minor variability. On the figure we show the profiles in 1999 and in 2006, both of the spectra are very high resolution, from ELODIE and ESPaDOnS, respectively.

Shortly before the dimming minimum, in August 2022 DAO spectrum, we see that all lines have shifted to higher radial velocities, have become significantly weaker than normal, and the line width has decreased. This can be attributed to the effect of the added blue-shifted emission that has appeared in the spectrum (strongly visible in H$\alpha$). For the \ion{Fe}{I} lines, the emission is not strong enough to reach higher than the continuum level. The emission strengthens and becomes even more apparent in the January 2023 FIES spectrum, where the resolution is sufficiently high to trace the small emission bump near $-70$~\kms. 

The added emission component in the \ion{Fe}{I} and many other absorption lines regrettably means that the line-depth-ratio method will be unreliable for estimating the temperature of RW\,Cep during the dimming period \citep[see longer discussion on the method applicability to YHGs in][]{kasikov_yellow_2024}. Therefore, we will rely on the $T_{\mathrm{eff}}$ measurements provided by \cite{Anugu2023} of 4200\,K when the star was in its calm state and 3900\,K during the dimming minimum. The variability in temperature is also supported by the change in (\textit{V-Ic}) colour index and the shift in its variability behaviour (Fig.~\ref{fig:V-Icorr}).

After the dimming minimum, by September 2023, the emission has almost disappeared in H$\alpha$, and we see this also in \ion{Fe}{I} lines. The two stronger \ion{Fe}{I} lines $\lambda\lambda6358,6430$ no longer have any indication of an emission affecting their blue wings. Additionally, they have become significantly stronger, even stronger than before the dimming. This could be caused by the decrease in $T_{\mathrm{eff}}$, which would strengthen lines with lower {\Elow} values (\ion{Fe}{I} $\lambda6358$ has \Elow=0.859 and $\lambda6430$ has \Elow=2.176). The weaker \ion{Fe}{I} lines $\lambda\lambda6322,6411$ show a slightly different behaviour. For these lines, we see the disappearance of the added emission component in the blue wing, too. By September 2023, the lines have grown increasingly wider, especially showing also an enhanced red absorption wing. However, the line profile has acquired an almost rectangular shape, as the lines have remained shallower than normal. The decreased $T_\mathrm{eff}$ during the dimming period could account for the shallower line depth in lines with higher \Elow, however, it does not account for the increased line width.

Flat-bottomed absorption lines have been recorded in the spectrum of $\rho$\,Cas. These are strong low-energy absorption lines, where the line profile is affected by added emission from cooler and diffuse shells near the star \citep{lobel_spectral_1998}. However, in our case, the weaker and higher {\Elow} lines have flat-bottomed profile shapes (\ion{Fe}{I} $\lambda6322$ has \Elow=2.588 and $\lambda6411$ has \Elow=3.654), so it would be reasonable to think that any persisting emission effect would primarily affect the lines with lower \Elow values. 

\begin{figure}
    \centering
    \includegraphics[width=\linewidth]{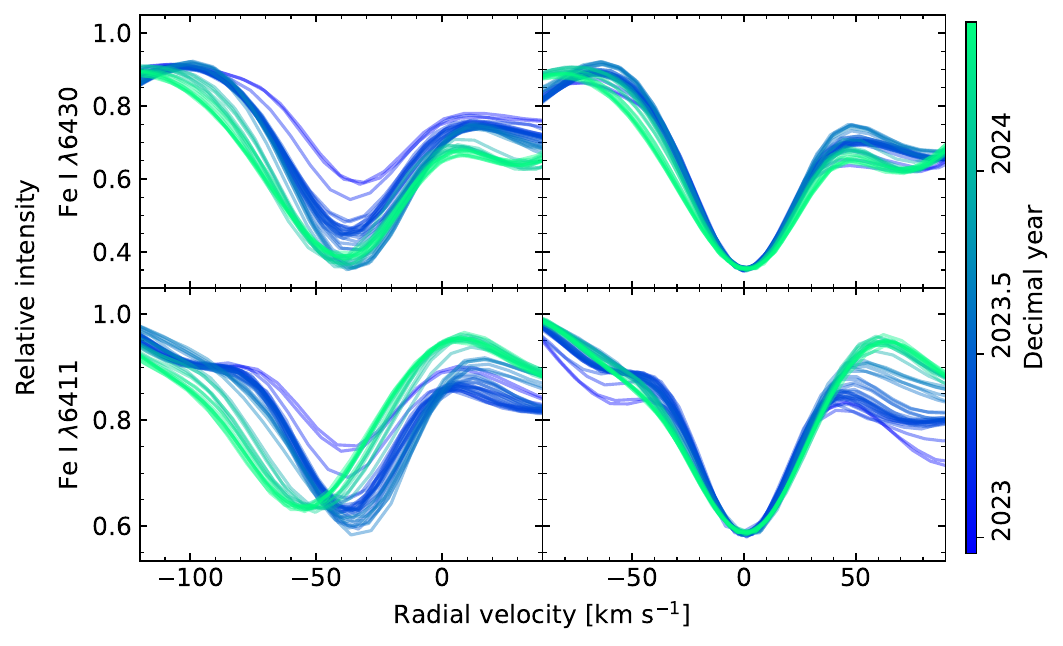}
    \caption{Variability of two \ion{Fe}{I} line profiles in TO spectra. Top row: $\lambda6430$, bottom row: $\lambda6411$. The left column shows the line profiles at their observed velocities, and the colour of the line marks time (from dark to light blue, starting from dimming minimum). In the right column, the same profiles have been scaled in core depth and shifted together in velocity direction. }
    \label{fig:to_stacked}
\end{figure}

\begin{figure}
    \centering
    \includegraphics[width=\linewidth]{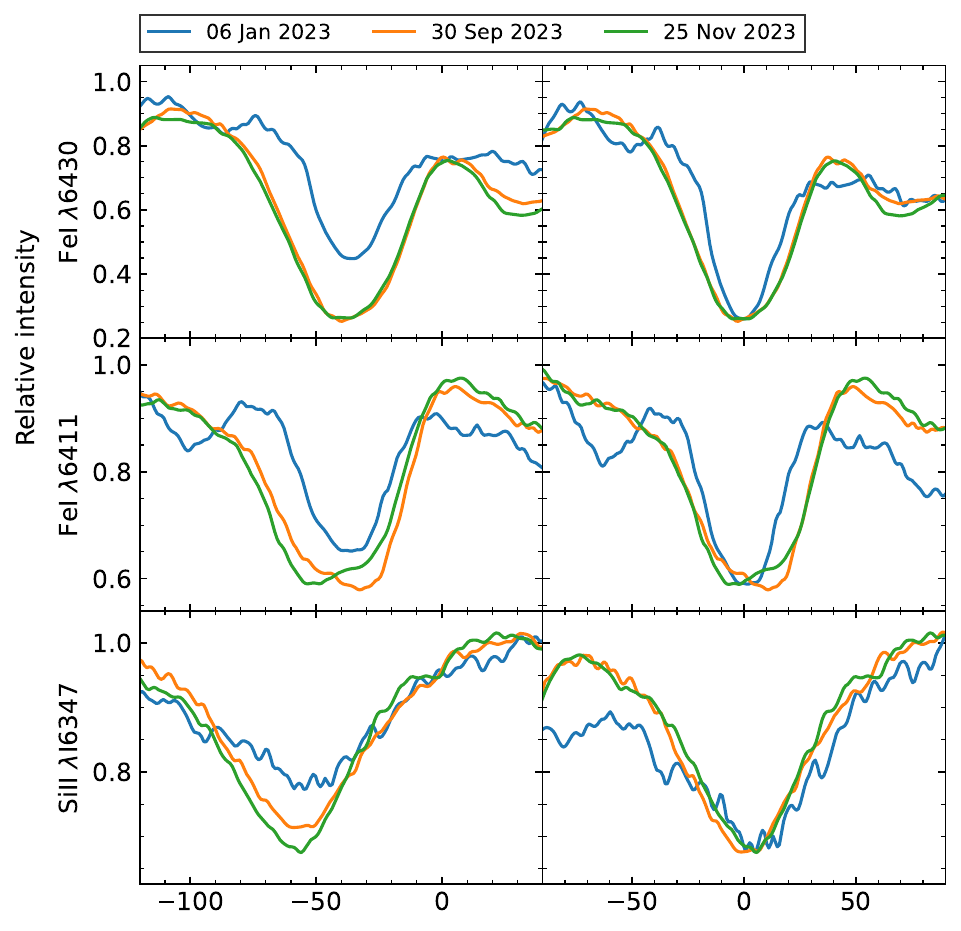}
    \caption{Same as Fig.\,\ref{fig:to_stacked} with addition of the \ion{Si}{II} line. We have plotted FIES spectra on three dates. The January spectrum (blue) is taken in dimming minimum, the other two are taken 9 months (orange) and 11 months (green) later, when the brightness had returned to near-normal level. }
    \label{fig:fies_stacked}
\end{figure}

Figure \ref{fig:to_stacked} illustrates the variability of a couple of line profiles during the Great Dimming \citep[following the example of Betelgeuse in][]{gray_mass_2008,gray_observation_2022}. The left column shows the line profiles of TO spectra on all epochs drawn on top of each other, the colour gradient traces the time axis -- darkest blue is during the dimming minimum and lightest green is in mid-2024. The right column shows the same line profiles, but centred in radial velocity and scaled to the depth of the deepest line. This exercise has been done for two \ion{Fe}{I} lines: $\lambda6430$ line in the top panels and $\lambda6411$ on the bottom panels. The \ion{Fe}{I} $\lambda6430$ is a very strong line that shows minor changes in its {\rv} during the event, up to 10\,\kms. While the \ion{Fe}{I}~$\lambda6411$ is a weaker line that shows significantly more variability in \rv, up to {25\,\kms} (c.f. the {\rv} curves in Fig.~\ref{fig:FeIradvel} and in App.~\ref{app:FeIradvels}). With the lower-resolution spectra of TO, {\rv} shifts dominate the variability behaviour, as is expected for a (hyper)giant star with giant convection cells that vary in timescales of months or years \citep{gray_observation_2022}. But we can also see from the scaled profiles that over time both lines grow slightly wider. 

The much higher resolution FIES spectra (Fig.\,\ref{fig:fies_stacked}) reveal the intricacies of the variability in the shape of the spectral lines. The FIES spectra show how the added emission component affects the line profile shape during the Great Dimming. We see that in January 2023 the line profiles are much shallower and narrower. When scaled to the same depth, the added emission contribution can be seen in both wings of the line. In the \ion{Si}{II} line, there is no indication of the line profile getting narrower, and neither does the radial velocity curve show a velocity maximum during the dimming minimum (Fig.~\ref{fig:FeI_radvel_zoom}). It is slightly shallower in January, which can be explained by the $T_{\mathrm{eff}}$ decrease. 

The September and November profiles are almost identical for the $\lambda6430$ line, but the weaker $\lambda6411$ line shows a change from a more enhanced red wing absorption to a stronger absorption in the blue wing. This change in line profile shape causes the radial velocity of this weaker line to decrease quite rapidly in comparison to the stronger lines (c.f. Fig.~\ref{fig:FeI_radvel_zoom}). The \ion{Si}{II} line shows a decrease in radial velocity, while the line depth increases.

\subsection{Absorption line splitting}

Back in 1956, \citet{merrill_complex_1956} found several strong resonance lines with two absorption components, separated by a central maximum (e.g. \ion{Ca}{I}, \ion{Cr}{I}, \ion{Mn}{I}, \ion{Sr}{II}, \ion{Ti}{I}). They marked that the lines were very asymmetrical and that the central maximum was very narrow, possibly originating from a superimposed emission. The mean velocity of the central maximum over 42 lines was at $-59.2$~\kms. \cite{gahm_spectral_1972} found a central emission component in \ion{Na}{I} D line doublet with a mean velocity of $-62$~\kms, in good agreement with \citet{merrill_complex_1956}, indicating that the physical processes behind the line profiles were similar in 1955 and 1969, when the individual authors took their observations. 

\citep{josselin_atmospheric_2007} studied the shapes of line profiles formed at different depths of the atmosphere in a sample of RSGs, including RW\,Cep. For RW\,Cep, they found that lines formed in the outermost layers of the atmosphere consist of two components: one of them systematically blue-shifted and the other red-shifted relative to the deeper layers. They attribute the line doubling to ascending and descending gas in a small number of giant convective cells. The velocity dispersion would cause the centre of the line profile to split. \citet{lobel_spectral_1998} have described a visually similar splitting effect in metallic absorption line cores in the spectrum of a YHG $\rho$\,Cas. In this case, the splitting occurs for both neutral and ionised atoms, and they link it to very low excitation energies (\Elow~<~1.6\,eV), and deep and saturated cores. However, for $\rho$\,Cas, the split profile shape is caused by the superposition of a static narrow emission upon a broad and strong absorption line. They have given as an example the line profile of the \ion{Ba}{II} $\lambda6141.321$ resonance line (\Elow = 0.7040\,eV). This gives us here an opportunity to draw a direct parallel between RW\,Cep and $\rho$\,Cas. 

\begin{figure}[h!]
    \centering
    \includegraphics[width=\linewidth]{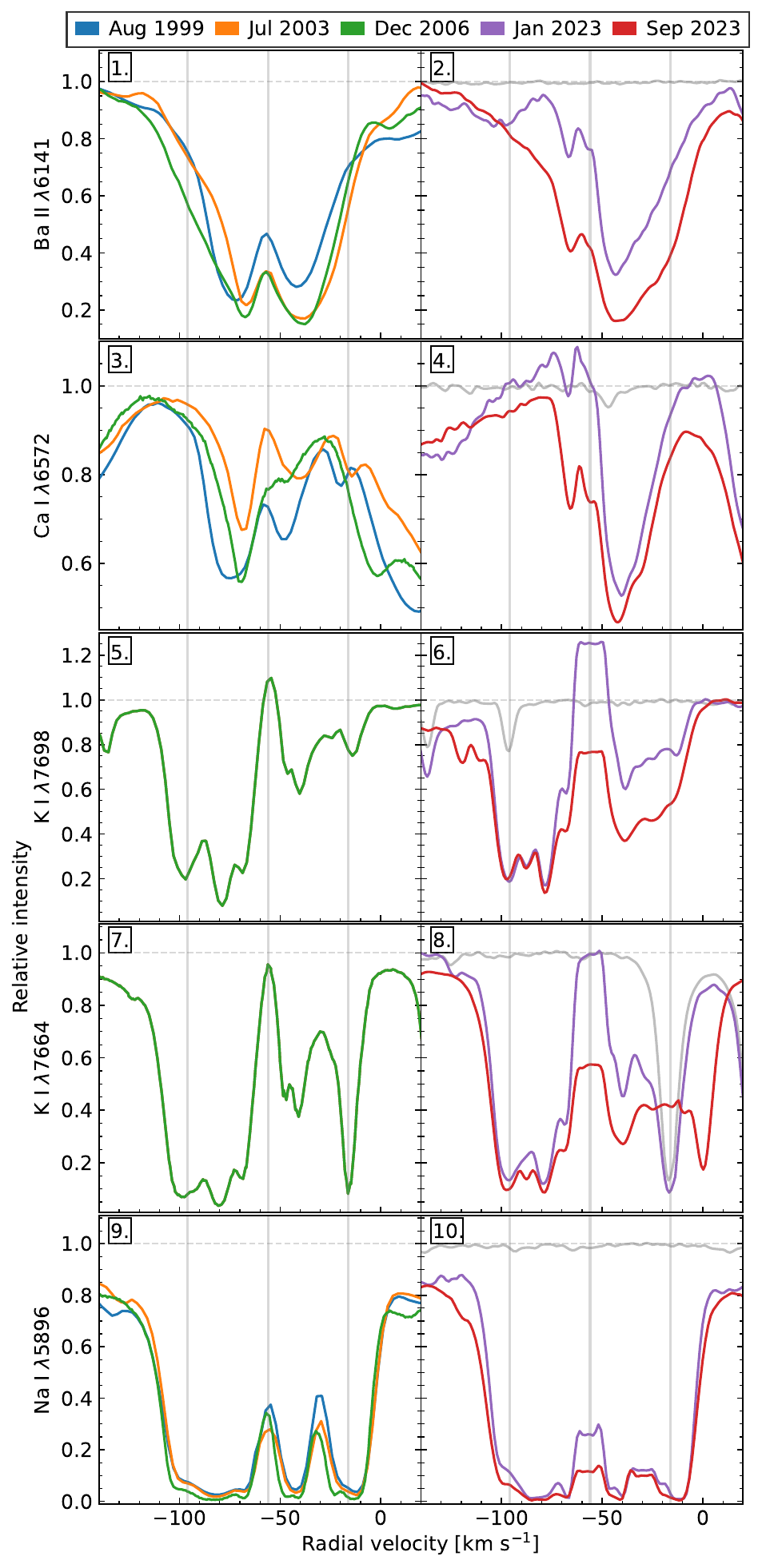}
    \caption{A selection of resonance lines in the spectrum of RW\,Cep. Rows from top to bottom: \ion{Ba}{II} $\lambda6141.71$, \ion{Ca}{I} $\lambda6572.78$, \ion{K}{I} doublet $\lambda\lambda7698.96,7664.91$ and \ion{Na}{I} D$_2$ $\lambda5895.92$. Odd-numbered panels show the line profiles in August 1999 (blue, ELODIE), July 2003 (orange, ELODIE) and December 2006 (green, ESPaDOnS) and even-numbered panels show the profiles during the Great Dimming minimum in January 2023 (purple, FIES) and 9 months later in September 2023 (red, FIES). Additionally, a telluric standard spectrum from the same January 2023 night has been included (light grey) to display the locations of blending telluric lines, especially in \ion{K}{I}. For the \ion{K}{I} only ESPaDOnS and FIES spectra are displayed due to the limited spectral range of ELODIE. Grey vertical lines trace the location of the emission peak at {$-56$~\kms} and also at {$\pm40$~\kms} from it.}
    \label{fig:low_energy_line_profiles}
\end{figure}

In Fig.\,\ref{fig:low_energy_line_profiles} panel 1, we show the \ion{Ba}{II} profile shape in the spectrum of RW\,Cep that looks quite similar to the profile seen in $\rho$\,Cas \citep[c.f. Fig.~13 in][]{lobel_spectral_1998}. Three different epochs are shown in panel 1: August 1999 (blue), July 2003 (orange) and December 2006 (green). The velocity of the central emission peak stays constant in time, while the intensities of the blue and red absorption wings vary. The spectrum from 2006 introduces an enhanced blue wing, which is not seen in other spectra from 1999--2005. This epoch corresponds to the lowest recorded values of \rv. In the spectrum of RW\,Cep, the other line of the doublet at $\lambda6496$ is also visible and has a very similar profile shape. 

In Fig.\,\ref{fig:low_energy_line_profiles} panel 3, we have drawn the \ion{Ca}{I} resonance line (\Elow = 0.0\,eV) profiles at same epochs. The \ion{Ca}{I} line shows a very similar two-bottom profile shape. The intensities of the absorption components vary more significantly than in \ion{Ba}{II}, the red component almost disappears in 2006, while the blue component strengthens. The emission peak in the \ion{Ca}{I} line has the same velocity as in \ion{Ba}{II} line, at approximately $-56$~\kms, where we have drawn a grey vertical line. We have included two more vertical lines at {$\pm40$\,\kms} from the emission peak velocity to help guide the eye. 

In the same figure, in panels 2 and 4 we show the \ion{Ba}{II} and the \ion{Ca}{I} line profiles during the Great Dimming minimum (January 2023, purple) and nine months later (September 2023, red). Additionally, we include a telluric standard spectrum taken during that same January night, plotted in light grey colour. During the January brightness minimum, the blue wings of both lines have almost completely disappeared, most likely due to the added blue-shifted emission affecting the lines at that time. The emission completely fills the line wing, leaving only a small nook near {$-60$~\kms} to mark the absorption component. The emission strength has decreased by September, when we see a deeper line profile. 

To explain the origin of the central emission component, we find it instructive to include the line profile of the \ion{K}{I} resonance doublet at $\lambda\lambda7664,7698$. The profiles before and during the dimming are displayed in Fig.~\ref{fig:low_energy_line_profiles} in panels 5--8. The profiles of these lines are complex, they are not blended with other metal lines, but some absorption components may have an interstellar origin. There are also notable telluric blends (see panels 6 and 8 where the telluric lines are drawn in grey for the January spectrum), at {$\sim-95$~\kms} in $\lambda$7698 line and at {$\sim-15$~\kms} in $\lambda7664$ line. Please note that the telluric line positions are slightly shifted in the September 2023 spectrum in relation to the January spectrum. 

Having noted the blending effects, let us focus on panels 5 and 7, which show the line profile in December 2006. Regrettably, the \ion{K}{I} lines are not in the spectral range of ELODIE. The central emission component in \ion{K}{I} is located at the same velocity as in the \ion{Ba}{II} and the \ion{Ca}{I} lines. And most importantly, the emission reaches approximately 10\% above the continuum level in the \ion{K}{I} $\lambda7698$ line. This excludes the possibility that the central emission feature appears due to a superposition of two absorption components with different velocities. Two absorption lines could not form a central feature that would rise above the continuum level. Additionally, the velocity of the emission components remains remarkably stable in time, even during the Great Dimming. This implies a similar emission origin, as in the case of $\rho$\,Cas -- a stable and cooler static envelope \citep{lobel_spectral_1998}. Interferometric observations show the vast atmosphere and historical light curves show evidence of multiple previous mass-loss outbursts \citep{anugu_time-evolution_2024,jones_recent_2023}. This creates an environment of cool and diffuse gas surrounding the star -- its upper atmospheric layers transition smoothly to circumstellar matter. Similar split line profiles of strong absorption lines that are composed of atmospheric and circumstellar components have been found in AGB stars \citep{klochkova_circumstellar_2014}. 

The absorption of \ion{K}{I} is a complex structure of multiple components that remain relatively stable in time. The blue wing is significantly wider than in \ion{Ba}{II} and \ion{Ca}{I} and it does not show major variability during the dimming period, in contrast to the former two lines. The red wing of the \ion{K}{I} lines behaves similarly to the other two lines. The deepest point of the red wing is located at {$\sim-40$~\kms} for all lines, and the wing deepens from January to September. The \ion{Na}{I} D lines (panels 9 and 10) have a very similar structure to the \ion{K}{I} lines. Both doublets have strongly saturated absorptions. It is likely because of that, we do not see the added emission affecting their blue wing during the dimming period. In addition to the added emission, which we see clearly in H$\alpha$ and other lines, there would be enhanced absorption in the cloud of ejected gas and dust. In these deep and saturated lines, absorption dominates. 

The evolution of the emission component in \ion{K}{I} and \ion{Na}{I} D lines during the Great Dimming is interesting (see panels 6, 8 and 10). The strengthened emission in \ion{K}{I} during the dimming was noticed also by \cite{Leadbeater2023}. Our high-resolution FIES spectra reveal that in addition to increasing in strength, the emission gained a flat-topped shape during the brightness minimum (January 2023). In the subsequent months, its strength decreased, but the shape remained the same (c.f. September 2023). Asymmetric emission shape during the dimming period can be seen in the \ion{Ba}{II} and the \ion{Ca}{I} lines as well. The shape of the central emission line forms in a complex interplay of the circumstellar absorption we see during the calm period and the added processes during the dimming: both emission and enhanced absorption. The combination of these would result in the flat-topped shape of the emission line, and the enhanced absorption would cause the emission to weaken by September compared to the calm period (as seen in \ion{K}{I} and \ion{Na}{I} D lines).

\section{Conclusions}

We present a high temporal cadence spectroscopic time series that covers the variability of RW\,Cep from the brightness minimum of the Great Dimming and the return to near-normal levels. We include a radial velocity analysis based on the variability of some representative \ion{Fe}{I} lines formed at different depths in the stellar atmosphere. We showcase some unusually shaped spectral lines (\ion{Ba}{II}, \ion{Ca}{I}, \ion{K}{I} and \ion{Na}{I}) and their implications to the circumstellar environment. We place the results in the context of the previous variability behaviour of RW\,Cep in 1999--2006, during a time when the star had stable brightness.

During the Great Dimming, the H$\alpha$ line gains a strong emission component that is displaced slightly towards the blue wing. It reaches its maximum strength during the dimming minimum and declines in intensity following that.

The atmosphere of RW\,Cep has a velocity gradient of $\sim$10--20~\kms. The gradient remains stable in time throughout 1999--2006 and increases during the Great Dimming. There is a correlation that stronger lines have higher {\rv} values and weaker lines have lower {\rv} values, which corresponds to the motions inside giant convective cells. We see that atmospheric stratification in radial velocities remains stable over all observations in the last 25 years. We measured the $v_{\mathrm{sys}}$ of the star at $-50.3$~\kms.

The radial velocity curve during the great dimming reveals different behaviour for spectral lines of different strengths: the strongest lines have a radial velocity maximum during the dimming minimum from December 2023 to January 2023 while the radial velocity of the weaker lines changes only slightly or not at all in the case of \ion{Si}{II}. The likely cause for this is the added emission, which is blue-shifted. This results in the absorption component shifting redwards to higher velocities. The \ion{Si}{II} line forms deeper in the atmosphere and thus is not affected by the emission or the radial velocity changes at that time. The decrease in the line depth of \ion{Si}{II} and weaker \ion{Fe}{I} lines could be caused by the $\sim300$~K decrease in $T_\mathrm{eff}$ during the Great Dimming. 

All of our studied spectral lines have a radial velocity maximum approximately six months after the dimming minimum, in May--June 2023, when their velocities are {$\sim$10\,\kms} higher than in the pre-dimming period. By then, the emission strength has greatly decreased (to below half of its maximum value based on H$\alpha$ profiles), so it is probably a minor influence on the radial velocity changes. This radial velocity variability shows large-scale motions in the atmosphere of the star. Lines that are formed deeper in the atmosphere show a faster return to normal radial velocity values, while the lines that are formed at higher atmospheric layers remain displaced for a longer period of time.

More information about the circumstellar environment of RW\,Cep is revealed by the central emission component that splits the cores of strong resonance lines (e.g. \ion{Ca}{I}, \ion{Ba}{II}, \ion{K}{I} and \ion{Na}{I} D lines). The above-continuum height of the emission peak in \ion{K}{I} line in December 2006 and its velocity stability over time imply its circumstellar origin. It is the only spectroscopic feature in the spectrum of RW\,Cep that holds a stable velocity over time, at $-56$~\kms.

It will be interesting to see, how RW\,Cep continues to develop after the outburst and how long, if at all, it will take to return to its normal variability behaviour. Moreover, if there will be changes in its variability, as was the case for Betelgeuse \citep{dupree_great_2022}. Continued monitoring of RW\,Cep will be highly beneficial to learning more about its atmospheric behaviour and to follow the evolution path of a budding YHG.

\begin{acknowledgements}

Based on observations made with the Nordic Optical Telescope, owned in collaboration by the University of Turku and Aarhus University, and operated jointly by Aarhus University, the University of Turku and the University of Oslo, representing Denmark, Finland and Norway, the University of Iceland and Stockholm University at the Observatorio del Roque de los Muchachos, La Palma, Spain, of the Instituto de Astrofisica de Canarias. We thank A. Casasbuenas for taking the spectrum.\\

Based on spectral data retrieved from the ELODIE archive at Observatoire de Haute-Provence (OHP).\\

Based on observations obtained at the Dominion Astrophysical Observatory, Herzberg Astronomy and Astrophysics Research Centre, National Research Council of Canada. \\

Based on observations made with the Perek 2~m telescope at Ond\v{r}ejov Observatory, operated by the Astronomical Institute of the Czech Academy of Sciences.\\

We acknowledge with thanks the variable star observations from the AAVSO International Database contributed by observers worldwide and used in this research.\\

This work has made use of data from the European Space Agency (ESA) mission
{\it Gaia} (\url{https://www.cosmos.esa.int/gaia}), processed by the {\it Gaia}
Data Processing and Analysis Consortium (DPAC,
\url{https://www.cosmos.esa.int/web/gaia/dpac/consortium}). Funding for the DPAC
has been provided by national institutions, in particular the institutions
participating in the {\it Gaia} Multilateral Agreement.\\

This project has received funding from the European Union's Horizon Europe research and innovation programme under grant agreement No. 101079231 (EXOHOST), and from the United Kingdom Research and Innovation (UKRI) Horizon Europe Guarantee Scheme (grant number 10051045). \\

This project has received funding from the European Union’s Framework Programme for Research and Innovation Horizon 2020 under the Marie Skłodowska-Curie Grant Agreement No. 823734.\\

This work was supported by the Estonian Research Council grant PRG 2159.

\end{acknowledgements}

\bibliographystyle{aa}
\bibliography{references.bib}

\begin{appendix} 
\onecolumn
\section{Selected \ion{Fe}{I} lines}
\begin{table}[h!]
\caption{Properties of the selected \ion{Fe}{I} lines. }
\label{table:FeIlines}
\centering  
\begin{tabular}{l c c c}
\hline\hline 
$\lambda$ [\AA] & \Elow [eV] & \loggf & Line depth\\
\hline
    6159.37* & 4.607 & -2.093 & 0.10 \\
    6180.20* & 2.7275 & -2.586 & 0.41 \\
    6226.73* & 3.883 & -2.220 & 0.17 \\
    6322.69 & 2.588 & -2.426 & 0.49\\
    6335.33 & 2.198 & -2.177 & 0.63\\
    6344.15 & 2.433 & -2.919 & 0.48\\
    6353.84 & 0.915 & -6.650 & 0.18\\
    6355.03 & 2.845 & -2.350 & 0.48\\
    6358.69 & 0.859 & -4.468 & 0.62\\
    6380.74 & 4.186 & -1.376 & 0.29\\
    6388.40* & 3.368 & -4.476 & 0.10 \\
    6393.60 & 2.433 & -1.432 & 0.62\\
    6408.02 & 3.686 & -1.230 & 0.45\\
    6411.65 & 3.654 & -0.595 & 0.50\\
    6419.94* & 4.733 & -0.240 & 0.38\\
    6421.35 & 2.279 & -2.027 & 0.64\\
    6430.84 & 2.176 & -2.006 & 0.68\\
    6469.19 & 4.835 & -0.770 & 0.27\\
    6475.62 & 2.559 & -2.942 & 0.43\\
    6546.24 & 2.758 & -1.536 & 0.52\\
    6569.21 & 4.733 & -0.420 & 0.34\\
    6597.56* & 4.795 & -2.070 & 0.16 \\
    6646.93* & 2.608 & -4.250 & 0.15 \\
    6648.08* & 1.011 & -6.175 & 0.25 \\
    6665.43* & 1.557 & -4.701 & 0.13 \\
    6667.42* & 2.453 & -4.549 & 0.16 \\
    6677.98* & 2.692 & -1.418 & 0.60\\
    6703.57* & 2.758 & -3.160 & 0.27 \\
    6705.10* & 4.6069 & -1.392 & 0.20 \\
    6710.32* & 1.485 & -5.050 & 0.29\\
    6726.67* & 4.607 & -1.133 & 0.22 \\
    6750.15* & 2.424 & -2.621 & 0.49\\
    6752.70* & 4.6380 & -1.204 & 0.20 \\
\hline 
\end{tabular}
\tablefoot{The line depth is measured from the continuum level to the lowest point of the line, a higher value indicates a stronger line. The given number is an average value calculated over all ELODIE spectra in the early 2000s. Line depth error is $\sim 0.006$ relative to the continuum. The lines marked with an asterisk (*) have not been measured from TO spectra because they are either out of the wavelength range or too heavily blended. }
\end{table}

\clearpage
\onecolumn
\section{Variability of H$\alpha$ line profile in 1999--2024}\label{app:halpha}
\begin{figure}[h]
\centering
   \includegraphics[width=17cm]{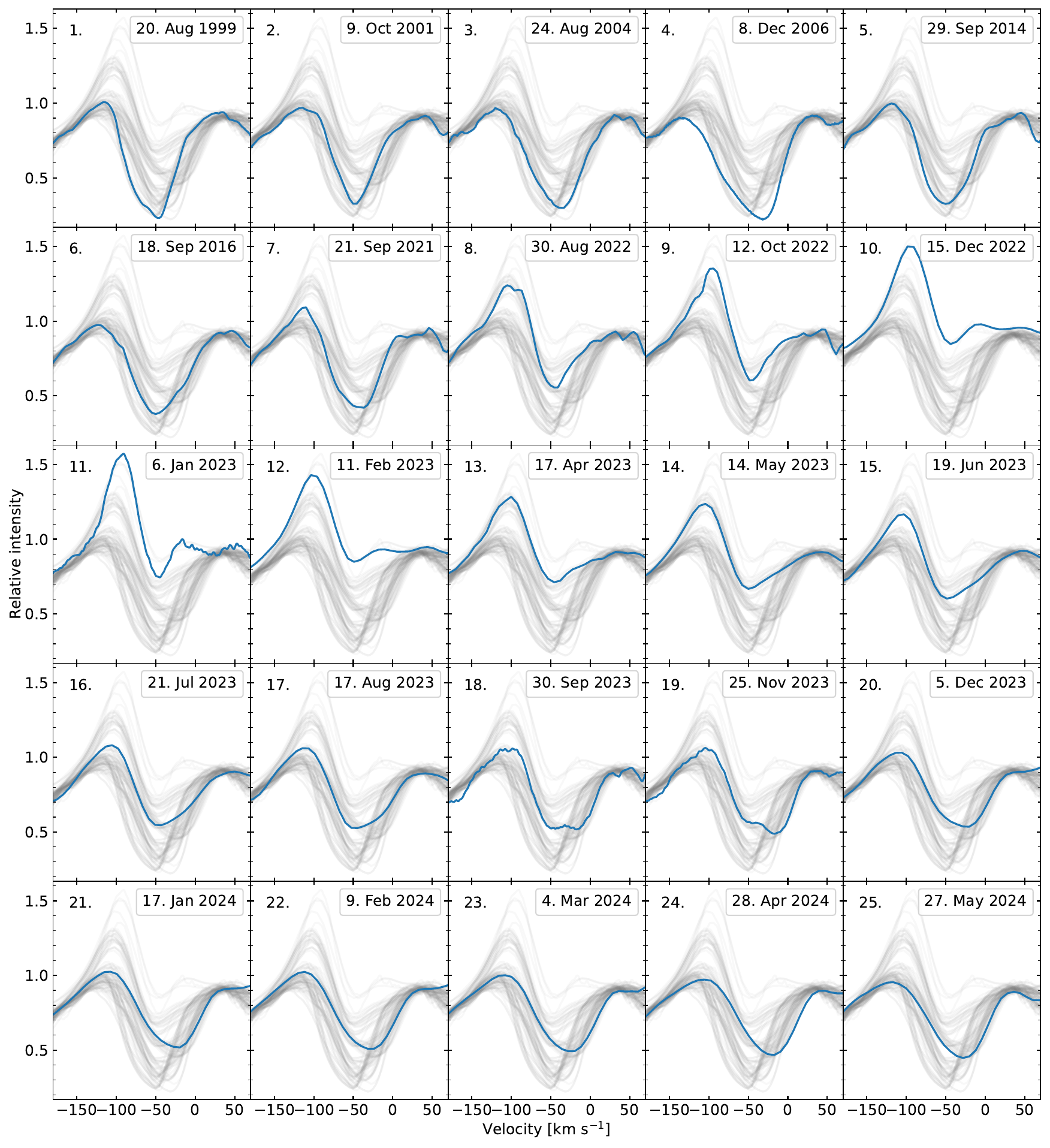}
     \caption{Selected line profiles to illustrate the variability of the H$\alpha$ profile before and during the Great Dimming. Please note that the systemic velocity has not been subtracted from the velocity axis. Individual dates are highlighted in blue and the thin grey lines in the background are all spectral profiles from all observation epochs, shown for comparison purposes. All observations are detailed in App. \ref{app:observations}. The spectra seem to form a couple of visually distinct 'groups', however, this is a side-effect of the uneven temporal coverage. Panels nr 1--6: H$\alpha$ line profile during the normal variability behaviour of the star. Panels 7--9: RW\,Cep is entering the dimming minimum, brightness is decreasing. Panels 10--11: Spectra taken during the dimming minimum, 15th December spectrum is from TO and has a lower resolution and 6th Jan is a high-resolution FIES spectrum. Panels 11--25: Post-dimming minimum evolution of the H$\alpha$ line. Spectra from dates Sept 2023 and Nov 2023 are from FIES, which reveal the finer structure in the H$\alpha$ line, rest of them are from TO.}
\end{figure}

\clearpage
\section{Radial velocity curves of \ion{Fe}{I}}\label{app:FeIradvels}
\begin{figure}[h]
    \centering
    \includegraphics[width=17cm]{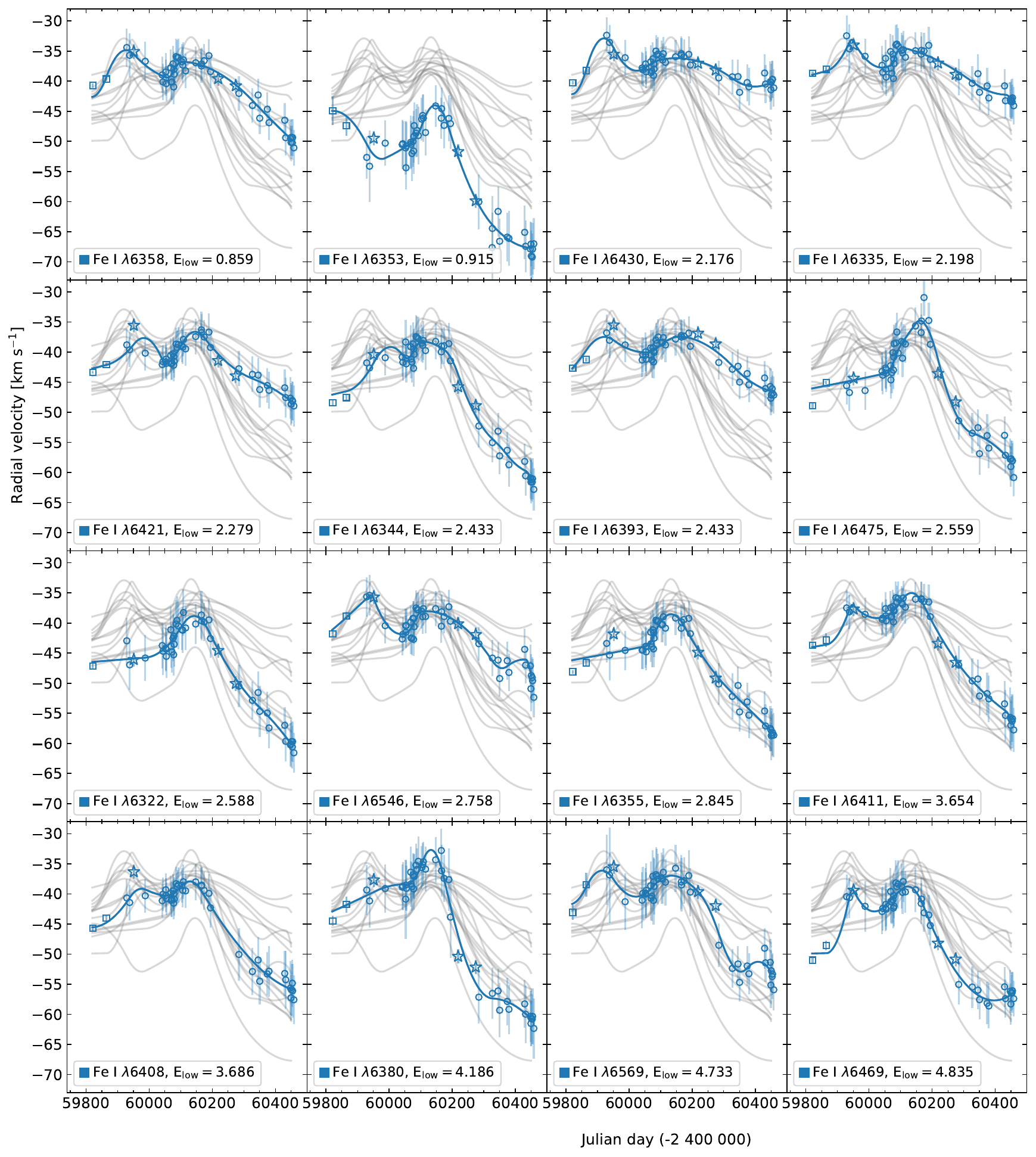}
    \caption{The radial velocity of 13 \ion{Fe}{I} lines during the Great Dimming episode. Each of the 13 panels highlights a different spectral line with the measured radial velocity data including uncertainties and a smooth trend line, the grey curves in the backround provide a point of reference towards the other lines. The shapes of the data points are the same as in the previous figures: square for DAO, star for FIES and circle for TO. The uncertainties of FIES and DAO data are smaller than the markers. The uncertainties of TO data are more significant (we estimate an additional error of {$\sim 2.5$\,\kms} not included in this plot). However, spectra taken in subsequent or nearby nights show good concurrence and also there is good match with FIES data. }
    \label{fig:FeIradveldetailed}
\end{figure}

\clearpage
\onecolumn
\section{Observations}
\begin{table}[hbt!]
\caption{Log of observations. }
\label{app:observations}
\centering
\begin{tabular}{l c c | l c c}
\hline\hline
Date & Observer & Instrument & Date & Observer & Instrument \\
\hline
31.07.1999 & Bennett & DAO & 18.09.2016 & Yang & DAO \\
01.08.1999 & Bennett & DAO & 21.09.2021 & Robotic & DAO \\
02.08.1999 & Bennett & DAO & 30.08.2022 & Robotic & DAO \\
03.08.1999 & Bennett & DAO & 12.10.2022 & Robotic & DAO \\
04.08.1999 & Yang & DAO & 15.12.2022 & Aret & TO \\
20.08.1999 & Josselin & ELODIE & 24.12.2022 & Checha & TO \\
15.09.1999 & Yang & DAO & 06.01.2023 & Kasikov & FIES \\
16.09.1999 & Yang & DAO & 11.02.2023 & Kasikov & TO \\
30.09.1999 & Yang & DAO & 06.04.2023 & Kasikov & TO \\
09.10.1999 & Yang & DAO & 08.04.2023 & Kasikov \& Borthakur & TO \\
10.10.1999 & Yang & DAO & 17.04.2023 & Mitrokhina & TO \\
25.02.2000 & Yang \& Brewster & DAO & 18.04.2023 & Borthakur & TO \\
27.06.2000 & Wallerstein & DAO & 19.04.2023 & Borthakur & TO \\
15.10.2000 & Yang & DAO & 03.05.2023 & Kasikov & TO \\
10.11.2000 & Yang & DAO & 04.05.2023 & Borthakur & TO \\
26.05.2001 & Yang & DAO & 07.05.2023 & Kasikov & TO \\
18.06.2001 & Yang & DAO & 10.05.2023 & Mitrokhina & TO \\
09.07.2001 & Yang & DAO & 12.05.2023 & Kasikov & TO \\
30.07.2001 & Yang & DAO & 13.05.2023 & Borthakur & TO \\
05.10.2001 & Yang & DAO & 14.05.2023 & Kasikov & TO \\
09.10.2001 & Yang & DAO & 20.05.2023 & Kasikov & TO \\
13.01.2002 & Yang & DAO & 23.05.2023 & Mitrokhina & TO \\
20.06.2002 & Stevi & DAO & 27.05.2023 & Mitrokhina & TO \\
21.06.2002 & Stevi & DAO & 07.06.2023 & Borthakur & TO \\
06.07.2002 & Yang & DAO & 10.06.2023 & Borthakur & TO \\
29.05.2003 & Josselin & ELODIE & 12.06.2023 & Kasikov & TO \\
04.07.2003 & Josselin & ELODIE & 19.06.2023 & Kasikov & TO \\
05.07.2003 & Josselin & ELODIE & 21.07.2023 & Checha & TO \\
07.07.2003 & Yang & DAO & 08.08.2023 & Kasikov & TO \\
20.08.2003 & Josselin & ELODIE & 09.08.2023 & Borthakur & TO \\
16.09.2003 & Josselin & ELODIE & 17.08.2023 & Aret & TO \\
14.01.2004 & Lebre & ELODIE & 01.09.2023 & Kasikov & TO \\
14.02.2004 & Josselin & ELODIE & 06.09.2023 & Kasikov & TO \\
27.04.2004 & Josselin & ELODIE & 30.09.2023 & Kasikov \& Checha & FIES \\
21.05.2004 & Josselin & ELODIE & 25.11.2023 & Casasbuenas & FIES \\
23.06.2004 & Josselin & ELODIE & 05.12.2023 & Checha & TO \\
24.07.2004 & Josselin & ELODIE & 16.01.2024 & Kasikov & TO \\
13.08.2004 & Be & DAO & 17.01.2024 & Checha & TO \\
24.08.2004 & Lebre & ELODIE & 04.02.2024 & Kasikov & TO \\
25.09.2004 & Yang & DAO & 09.02.2024 & Checha & TO \\
02.02.2005 & Quidam & ELODIE & 04.03.2024 & Eenm\"ae & TO \\
09.04.2005 & Yang & DAO & 09.03.2024 & Kasikov & TO \\
09.08.2005 & Yang & DAO & 28.04.2024 & Kasikov & TO \\
23.09.2005 & Yang & DAO & 01.05.2024 & Kasikov & TO \\
08.10.2005 & Yang & DAO & 16.05.2024 & Borthakur & TO \\
25.09.2006 & Yang & DAO & 18.05.2024 & Kasikov & TO \\
08.12.2006 & Lebre & ESPaDOnS & 19.05.2024 & Mitrokhina & TO \\
09.12.2006 & Lebre & ESPaDOnS & 22.05.2024 & Borthakur & TO \\
05.09.2011 & Aret & Ond\v{r}ejov & 23.05.2024 & Mitrokhina & TO \\
29.09.2014 & Yang & DAO & 27.05.2024 & Kasikov & TO \\
\hline
\end{tabular}
\tablefoot{The ELODIE spectra are available in the ELODIE archive (\url{http://atlas.obs-hp.fr/elodie/}); the DAO spectra are available in the Canadian Astronomy Data Centre (\url{https://www.cadc-ccda.hia-iha.nrc-cnrc.gc.ca/}); ESPaDOnS spectra are available from Polarbase (\url{http://polarbase.irap.omp.eu/}); FIES spectral data are available in the Nordic Optical Telescope arhcive (\url{https://www.not.iac.es/archive/}); Ond\v{r}ejov and TO data are available upon request from the corresponding author.}
\end{table}

\end{appendix}

\end{document}